\documentclass[a4paper,12pt]{article}

\usepackage{soul}
\usepackage[usenames,dvipsnames]{color}
\usepackage{jheppub}
\usepackage{amsmath, amssymb, slashed, epsf, color, graphicx, latexsym}
\usepackage{epsfig}

\usepackage{color}

\begin{document}
	\preprint{}
	\title{Black Rings in Large $D$ Membrane Paradigm at the First Order}

	\author{Mangesh Mandlik}
	
	\affiliation{Department of Physics, Indian Institute of Technology, Kanpur 208016, India.}

	\emailAdd{mangeshm@iitk.ac.in}

	\abstract{Black rings are the black objects found in $D$ spacetime dimensional gravity when $D \geq 5$. These have event horizon topology $S^{D-3}\times S^1$. In this work the solutions of the large $D$ membrane paradigm dual to stationary black rings in Einstein-Maxwell theory with or without cosmological constant are studied. It is shown that the first order membrane equations can only admit static asymptotically flat black rings, and the equilibrium angular velocity for the asymptotically AdS black rings at large $D$ was obtained. The thermodynamic and dynamic stability of the asymptotically flat black ring solutions is studied. The apparent shortcomings of some of these results are argued to be curable within the large $D$ membrane paradigm framework.}
	
	\maketitle
	
\section{Introduction}\label{intro}

Einstein, in his infinite wisdom, formulated his famous theory of General Relativity to describe gravity in {\it four} spacetime dimensions, because that's how many he saw around him. Although the equations of motion defining this theory are coupled nonlinear differential equations of metric components, some very interesting features of this theory have been observed over years. One such feature is the existence of a class of solutions called `black holes'. A black hole solution has an `event horizon' which separates the spacetime into causally disconnected pieces. Years of analysis has shown that these black holes in $4$ dimensions have some universal characteristics. All the exact stationary black hole solutions in $D=4$ that have been found analytically, such as Schwarzschild, Kerr and Reissner-Nordstrom, have the event horizon with every `time slice' having the topology $S^2$, so people wondered if the dynamical black holes have some constraints on their topology. In fact, it was proved (see \cite{Hawking:1973uf}) that for all asymptotically flat black holes in $D=4$, each connected component of a time slice of the event horizon has the topology $S^2$. Other features include the `no-hair theorem' which states that every stationary black hole is identified only by its mass, angular momentum and $U(1)$ charge, the `uniqueness theorem' which states that for each combination of these parameters there is a unique stationary black hole, and the lack of perturbative instability for all stationary black holes.

With the advent of String Theory, especially since the proposal of AdS/CFT correspondence, there has been and increased interest in the black hole solutions in $D>4$. Some attempts have been made to see if the higher dimensional black holes have any constraints on their topology like their $4D$ counterparts. \cite{Cai:2001su} showed using 'topological censorship' that in asymptotically flat $D=5$ the topology of a time slice of a black hole event horizon must be a connected sum of $S^3$ (with some identifications) and $S^2 \times S^1$. The solutions with $S^3$ topology have been known for a long time: The higher dimensional versions of Schwarzschild (Schwarzschild - Tangherlini), Reissner - Nordstrom, and higher dimensional generalization of Kerr solution namely the Myers-Perry black holes \cite{Myers:1986un, Myers:2011yc} to name the ones which are in fact known for all $D \geq 4$. For the first time, Emparan and Reall \cite{Emparan:2001wk, Emparan:2001wn} constructed the exact asymptotically flat black hole solutions with $S^2 \times S^1$ horizon topology and named them 'black rings'. In \cite{Emparan:2006mm} the phase structure of Myers-Perry black holes and black rings was obtained (see fig.(2) in there) and it was shown that for a range of angular momenta the black hole and black ring coexist, showing that the uniqueness theorem doesn't hold in $5D$. Also, unlike the $4D$ black holes, these black rings are perturbatively unstable. So most of the features of $4D$ black holes don't carry over to higher dimensions.

For $D>5$, exact black ring solutions, which are defined to have $S^{D-3} \times S^1$ topology, can't be constructed by any known technique \cite{Chervonyi:2015uua}, so one needs to resort to approximations. A construction in \cite{Emparan:2007wm, Caldarelli:2008pz} used 'thin ring' approximation to get black rings with small angular momentum.  But beyond the validity of this approximation the solution loses accuracy, and hence a more complete phase diagram for $D>5$ black rings is only obtained numerically. The numerical analysis of $D=6$ \cite{Kleihaus:2012xh} shows that some features of the $5D$ black ring, like the cusp in fig.(2) of \cite{Emparan:2006mm} do exist in higher dimensions. It would be very useful to find the black ring solutions in other approximations in order to obtain a more complete picture of their phase structure.

One such approximation is made by taking $D$ to be very large, and then perturbatively correcting the solution in orders of $1/D$. There have been two parallel developments in finding the black hole solutions in general\footnote{See \cite{Emparan:2020inr} for a short but excellent review.}. One branch is led by Emparan, Suzuki, Tanabe (EST) and collaborators \cite{Emparan:2013moa, Emparan:2013xia, Emparan:2013oza, Emparan:2014cia, Emparan:2014jca, Emparan:2014aba, Emparan:2015rva, Emparan:2015hwa, Suzuki:2015iha, Suzuki:2015axa, Emparan:2015gva, Tanabe:2015hda, Tanabe:2015isb, Andrade:2015hpa, Emparan:2016sjk, Tanabe:2016pjr, Tanabe:2016opw, Andrade:2018zeb, Andrade:2018nsz, Emparan:2019obu, Andrade:2019edf, Licht:2020odx}, which has been very successful in finding various stationary black holes and studying the dynamics of instabilities\footnote{See \cite{Chen:2015fuf, Chen:2016fuy, Chen:2017wpf, Chen:2017hwm, Chen:2017rxa, Chen:2018nbh, Chen:2018vbv, Li:2019bqc, Guo:2019pte}, \cite{Herzog:2016hob, Rozali:2016yhw, Rozali:2017bll, Herzog:2017qwp, Rozali:2018yrv, Casalderrey-Solana:2018uag}, and \cite{Sadhu:2016ynd, Sadhu:2018zyh, Sadhu:2018asi} for other collaborations that use large $D$ techniques similar to the framework of EST to study interesting and diverse problems.}. In this framework, \cite{Tanabe:2015hda, Tanabe:2016pjr} and \cite{Chen:2017wpf} have produced respectively the uncharged and charged black ring solutions in large $D$ and studied their stability using quasinormal mode analysis.

While the other branch, known as the ``large $D$ membrane paradigm'' \cite{Bhattacharyya:2015dva, Bhattacharyya:2015fdk, Dandekar:2016fvw, Dandekar:2016jrp, Bhattacharyya:2016nhn, Bhattacharyya:2017hpj,Dandekar:2017aiv, Bhattacharyya:2018szu, Mandlik:2018wnw, Saha:2018elg, Kundu:2018dvx, Bhattacharyya:2018iwt, Bhattacharyya:2019mbz, Kar:2019kyz, Dandekar:2019hyc, Biswas:2019xip, Patra:2019hlq} states that the dynamical black holes in large $D$ have one-to-one correspondence to the dynamics of a codimension one hypersurface, called a ``membrane'', which doesn't backreact on its background. In this nongravitational dual, the membrane dynamics is governed by what are called the ``membrane equations of motion'', and  the dynamical data associated with the membrane completely defines the corresponding black hole solution. The membrane equations and the correspondence between two pictures can be systematically improved order by order in $1/D$. This scheme is potentially very powerful in studying the dynamical black holes that cannot be approximated by stationary ones. But its ability to produce stationary solutions as well was shown in \cite{Mandlik:2018wnw}. In fact as an example, along with membranes corresponding to rotating charged and uncharged black holes with or without cosmological constant, the membranes dual to asymptotically flat black rings were also obtained there.

The goal of this paper is to check which membrane configurations dual to black rings can be obtained from the first order large $D$ membrane paradigm, and study their properties like thermodynamics and perturbative stability. Structure of this paper is as follows. In section \ref{exist} the membrane equations are analysed to the first order to assess which class of such solutions is obtainable at this order. In section \ref{thermosec}, the thermodynamic stability of this black ring membrane is obtained by comparing it with a black hole membrane competing for the same spot in the parameter space. In section \ref{lightQNM}, the dynamical stability of the black ring is studied by finding the quasinormal fluctuation modes from the linearized membrane equations. The subsection at the end of each of these sections is dedicated to comparing the results of the respective section to the results in the existing literature, and to propose the ways to correct the results in this paper while staying in the membrane paradigm. The section \ref{theend} concludes the main body of the paper by summarizing the results and proposing future directions. In the end, the details of various calculations are given in the appendix.

A comment about the nomenclature used in this paper is in order. As mentioned above, the large $D$ membrane paradigm and the membrane equations deal with the membrane configurations in the nongravitational picture. But these configurations correspond to black objects in the gravitational picture. So when a ``black hole'' or a ``black ring'' is mentioned in the membrane paradigm context, the author is implicitly referring to the membrane configurations corresponding to the respective gravitational objects. For example, ``the subleading correction to the black ring'' means the subleading correction to the membrane dual to the black ring.

\section{Existence of black ring type stationary solutions}\label{exist}

In \cite{Mandlik:2018wnw} the authors specialised the first order membrane equations from the charged membrane paradigm with nonzero cosmological constant \cite{Kundu:2018dvx} to stationary cases, and explicitly found out some axisymmetric stationary solutions. While they could find black hole solutions (the ones with $S^{D-2}$ topology) for charged and uncharged membranes in flat and in (A)dS backgrounds, the black ring solutions (the ones with $S^{D-3} \times S^1$ topology) were found only in the flat background and they were static. However these solutions were found out with the help of a ``quadratic ansatz''. In this section we will see that the form of the first order stationary membrane equations themselves restricts the possible types of axisymmetric stationary solutions.

\subsection{Charged black rings in AdS}\label{BRlead}
In \cite{Mandlik:2018wnw}, axisymmetric membrane solutions were obtained by solving the stationary membrane equation in the $(t,r,\theta,s,\{\chi^a\})$ coordinate system. The background (A)dS space metric in these coordinates is written as
\begin{equation}
-d\tau^2 = -dt^2\left(1+\frac{r^2+s^2}{L^2}\right) + dr^2 + ds^2 - \frac{(rdr+sds)^2}{r^2+s^2+L^2} + r^2d\theta^2 + s^2 d\Omega^2_{D-4}. 
\end{equation}
Where $L$ is the (A)dS radius and $d\Omega^2_{D-4}$ is the volume element of $S^{D-4}$ spanned by the $(D-4)$ angular coordinates $\{\chi^a\}$. For de Sitter, $L^2$ is analytically continued to negative values, while in $L \to \infty$ limit the metric becomes flat. $\theta$ is the angle in the polar coordinates $(r,\theta)$ describing the separated out `rotation plane'.

The symmetries of the problem (time independence due to stationarity, spherical symmetry in $S^{D-4}$ directions and $\theta$ independence due to axisymmetry) dictate that a membrane obeying these symmetries can be written in the form 
$$s^2 = 2g(r),$$
whereas the stationary membrane equations
\begin{equation}\label{stMeq}
\begin{split}
K &= \frac{4\pi T\gamma}{1-Q^2},\\
Q &= 2\sqrt{2\pi}\mu\gamma,
\end{split}
\end{equation}
which together become
\begin{equation}
K = \frac{4\pi T\gamma}{1-8\pi\mu^2\gamma^2},
\end{equation}
take the form of an ordinary differential equation for the function $g(r)$ when we choose $u = \gamma(\partial_t + \omega\partial_\theta)$. $\gamma$ is the normalization chosen such that $u\cdot u = -1$. Solving this ODE gives the shape of the membrane configuration dual to an axisymmetric stationary solution.

We borrow this ODE obtained from the stationary membrane equations in the $(t,r,\theta,s,\{\chi^a\})$ coordinates from \cite{Mandlik:2018wnw} (eq. [4.63] in it) and specialise it to the singly rotating case:
\begin{equation}\label{ChAdS}
\begin{split}
&\left(2g + \left(\frac{dg}{dr}\right)^2 + \frac{1}{L^2}\left(2g-r\frac{dg}{dr}\right)^2\right)\left(1+\frac{2g}{L^2}-r^2\left(\omega^2-\frac{1}{L^2}\right)\right)\\
=& \beta^2\left(1+\frac{1}{L^2}\left(2g-r\frac{dg}{dr}\right)\right)^2\left(1+\frac{2g}{L^2}-r^2\left(\omega^2-\frac{1}{L^2}\right)-\alpha^2\right)^2,
\end{split}
\end{equation}
where parameters $\alpha \equiv 2\sqrt{2\pi}\mu$ and $\beta \equiv \frac{D}{4\pi T}$ are defined to simplify the equation.$\alpha$ is the charge parameter; when we set it to zero we explore the uncharged stationary solutions. Now we make all the quantities dimensionless by defining
\begin{equation}
h\equiv \frac{g}{\beta^2},~~~x\equiv\frac{r}{\beta},~~~l\equiv\frac{L}{\beta},~~~\Upsilon\equiv\beta\omega,~~~~~'\equiv \frac{d}{dx},
\end{equation}
which turns \eqref{ChAdS} into
\begin{equation}\label{ChAdSdl}
\begin{split}
&\left(2h + (h')^2 + \frac{\left(2h-xh'\right)^2}{l^2}\right)\left(1+\frac{2h}{l^2}-x^2\left(\Upsilon^2-\frac{1}{l^2}\right)\right)\\
=& \left(1+\frac{\left(2h-xh'\right)}{l^2}\right)^2\left(1+\frac{2h}{l^2}-x^2\left(\Upsilon^2-\frac{1}{l^2}\right)-\alpha^2\right)^2.
\end{split}
\end{equation}
Now let's investigate one of the extrema of the solution, $x=x_0$, and let $h_0\equiv h(x_0)$. Thus $h'(x_0) = 0$. At $x_0$, \eqref{ChAdSdl} becomes
\begin{equation}\label{ChAdSdl0}
2h_0\left(1+\frac{2h_0}{l^2}-x_0^2\left(\Upsilon^2-\frac{1}{l^2}\right)\right)
= \left(1+\frac{2h_0}{l^2}\right)\left(1+\frac{2h_0}{l^2}-x_0^2\left(\Upsilon^2-\frac{1}{l^2}\right)-\alpha^2\right)^2.
\end{equation}
Here we have cancelled a factor of $1+\frac{2h_0}{l^2}$ from both sides. We expect it to be nonzero (and positive) because $\frac{2h_0}{l^2}\equiv\frac{s_{max}^2}{L^2}$ has to be nonnegative for the AdS and flat cases, while in the de Sitter case $1-\frac{2h_0}{|l^2|}=0$ means $\gamma^{-2} = 1-\frac{2h}{|l^2|}-x^2\left(\Upsilon^2+\frac{1}{|l^2|}\right)$ becomes nonpositive at $x=x_0$ which is illegal.\\
Now we differentiate \eqref{ChAdSdl} with $x$ and again evaluate it at $x_0$
\begin{equation}\label{ChAdSdldiff}
\begin{split}
&-\frac{4x_0h_0}{l^2}h_0''\left(1+\frac{2h_0}{l^2}-x_0^2\left(\Upsilon^2-\frac{1}{l^2}\right)\right)-4x_0h_0\left(1+\frac{2h_0}{l^2}\right)\left(\Upsilon^2-\frac{1}{l^2}\right)\\
=& -\frac{2x_0}{l^2}h_0''\left(1+\frac{2h_0}{l^2}\right)\left(1+\frac{2h_0}{l^2}-x_0^2\left(\Upsilon^2-\frac{1}{l^2}\right)-\alpha^2\right)^2\\
&-4x_0\left(\Upsilon^2-\frac{1}{l^2}\right)\left(1+\frac{2h_0}{l^2}\right)^2\left(1+\frac{2h_0}{l^2}-x_0^2\left(\Upsilon^2-\frac{1}{l^2}\right)-\alpha^2\right).
\end{split}
\end{equation}
Using \eqref{ChAdSdl0}, we notice that the terms involving $h_0''$ cancel out. Also, $x_0 = 0$ trivially solves \eqref{ChAdSdldiff}, which corresponds to the black hole solution, which has the maximum\footnote{Solving \eqref{ChAdSdldiff} indicates an extremum, but not necessarily a maximum. However, since we are demanding compact solution, $h=0$ at some finite $x=x_{max}$. As $h\geq 0$, if $x_0 = 0$ were a minimum, for a nontrivial solution there must be a maximum between $x=0$ and $x=x_{max}$. Then we shift our focus to that maximum instead.} in $h$ at $x=0$. But for other extrema, we choose $x_0 \neq 0$. Also, for the case $l^2\Upsilon^2 \neq 1$ (i.e. $\omega^2L^2 \neq 1$), \eqref{ChAdSdldiff} simplifies to
\begin{equation}\label{ChAdSdl1}
h_0 = \left(1+\frac{2h_0}{l^2}\right)\left(1+\frac{2h_0}{l^2}-x_0^2\left(\Upsilon^2-\frac{1}{l^2}\right)-\alpha^2\right).
\end{equation}
Comparing \eqref{ChAdSdl0} and \eqref{ChAdSdl1}, we get
\begin{equation}
1+\frac{2h_0}{l^2}-x_0^2\left(\Upsilon^2-\frac{1}{l^2}\right)-\alpha^2 = 2\left(1+\frac{2h_0}{l^2}-x_0^2\left(\Upsilon^2-\frac{1}{l^2}\right)\right),
\end{equation}
therefore,
\begin{equation}
1+\frac{2h_0}{l^2}-x_0^2\left(\Upsilon^2-\frac{1}{l^2}\right) = -\alpha^2.
\end{equation}
Plugging this back in \eqref{ChAdSdl1}.
\begin{equation}
h_0= -2\alpha^2\left(1+\frac{2h_0}{l^2}\right),
\end{equation}
which gives an absurd result that $h_0$ is nonpositive. Hence for the charged case ($\alpha \neq 0$), there can't be an extremum at any value of $x$ (i.e. $r$) other than zero. In particular, there is no black ring solution to the first order membrane equation.

In the uncharged case, $h_0=0$. But assuming there is a black ring solution, this means that $h=0$ at the endpoints as well as at all the extrema, thus giving $h=0$ everywhere in between. So this is also an absurd solution. Hence there is no black ring solution in the uncharged case as well. This means there are no stationary asymptotically (A)dS black ring solutions to the first order large $D$ membrane paradigm for $\omega^2 \neq \frac{1}{L^2}$.

The membranes in flat background can be investigated for black ring solutions by taking $l \to \infty$ and all the above steps go through provided $\Upsilon \neq 0$. This results into the proof that there is no rotating charged or uncharged black ring solution to the first order membrane equations in flat background either.

The above proof breaks down, however, for $\Upsilon^2 = \frac{1}{l^2}$. That is because the procedure to get \eqref{ChAdSdl1} from \eqref{ChAdSdldiff} involves dividing both sides by $\Upsilon^2-\frac{1}{l^2}$, which cannot be done any longer. $\Upsilon^2 = \frac{1}{l^2}$ can not ever hold for de Sitter case where $l^2 < 0$, so even this exception is not applicable to de Sitter case and thus the first order membrane paradigm doesn't have asymptotically de Sitter stationary black rings. In the static, flat case $\Upsilon = 0$ and $1/l = 0$. Also in AdS, $l^2 > 0$, and this breakdown corresponds to a particular value of angular velocity, $\omega = \frac{1}L$. These two cases will be discussed separately in the next two subsections.

\subsection{Static black ring in flat background}\label{flatring}
Let's look back at \eqref{ChAdSdl}, and let's put $\frac{1}{l} = \Upsilon = 0$. On simplification we get
\begin{equation}\label{Chfldl}
2h + (h')^2 = \left(1-\alpha^2\right)^2,
\end{equation}
Which is the dimensionless equation of static membrane in flat background in the $(t,r,\theta,s,\{\chi^a\})$ coordinates. At the extremum, it gives
\begin{equation}
2h_0 = \left(1-\alpha^2\right)^2.
\end{equation}
But differentiating \eqref{Chfldl} and evaluating at $x = x_0$ gives a null equation. So no more information can be fetched form this equation. However, \eqref{Chfldl} can be solved analytically by rearranging into
\begin{equation}
h' = \pm\sqrt{1-\alpha^2-2h},
\end{equation}
to give
\begin{equation}
2h = 1-\alpha^2-(x-c)^2,
\end{equation}
When the integration constant $c$ is set to zero, this is the Reissner-Nordstrom black hole solution. For $c\leq 1-\alpha^2$ the point $x=0$ is in the domain of the solution. For the solution to be regular at $x=0$, we demand $h'(0)=0$, which yields the condition $c=0$, i.e. the black hole solution. However, when $c > 1-\alpha^2$ the solution extends from $x=c-(1-\alpha^2)$ to $x=c+(1-\alpha^2)$, and has a maximum at $x=c$. The point $x=0$ is not a part of the membrane anymore, and so the regularity at $x=0$ is not required, and $c$ can take any value $\geq 1-\alpha^2$. This is our desired black ring solution. Since this solution is valid for any $\alpha$, the static black ring solution exists for both charged and uncharged membranes in flat background. Rescaling back, and recalling $2h\beta^2=2g = s^2$, the black ring solution in $(t,r,\theta,s,\{\chi^a\})$ coordinates is given by
\begin{equation}\label{brsol}
 s^2 + (r-b)^2 = R^2.
\end{equation}
\begin{figure}[!tbp]
		\centering
		\begin{minipage}[b]{0.48\textwidth}
		\includegraphics[width=\textwidth]{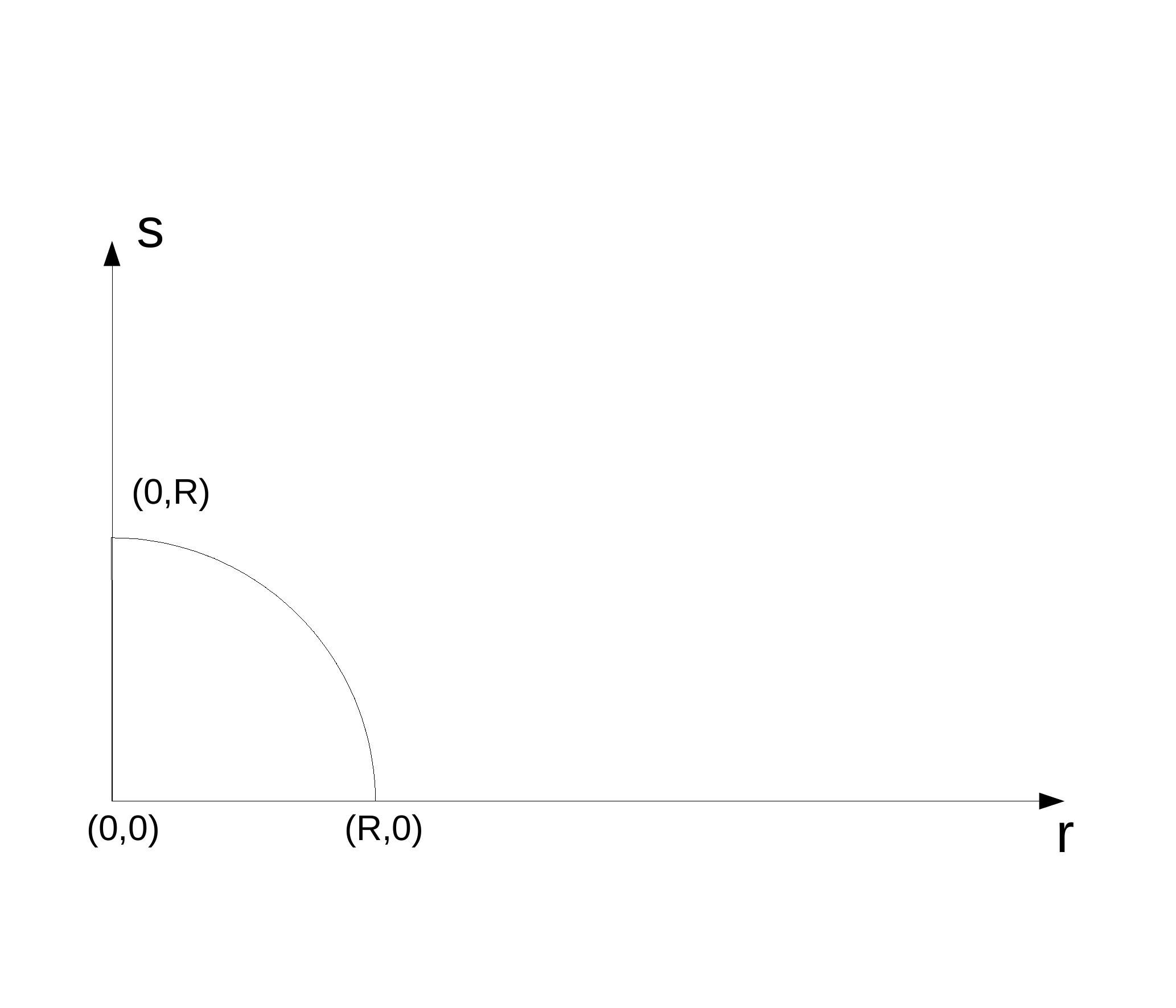}
		\caption{$s$ vs $r$ plot for a static black hole with radius $R$ ($b=0$).}
		\label{BHfig}
	\end{minipage}
		\hfill
		\begin{minipage}[b]{0.48\textwidth}
		\includegraphics[width=\textwidth]{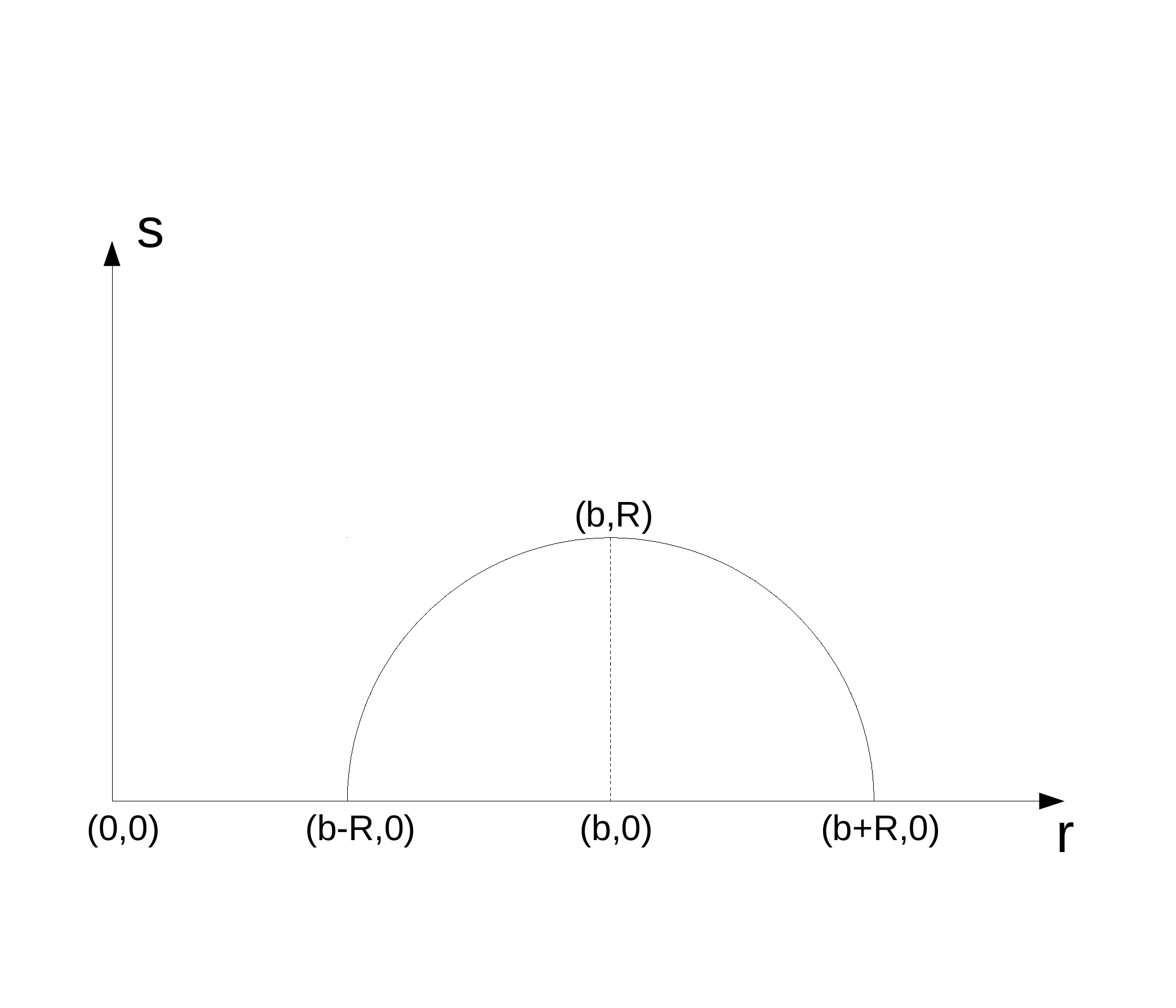}
		\caption{$s$ vs $r$ plot for a black ring with $S^{D-3}$ radius $R$ ($b>R$).}
			\label{BRfig}
		\end{minipage}	
\end{figure}\newpage
Where $b = \beta c$ is the $S^1$ radius (also called the ``ring radius'') while $R \equiv \beta(1-\alpha^2)$ is the $S^{D-3}$ radius of the ring. We always have $b > R$ according to the arguments in the previous paragraph\footnote{The limit $b \to R$ from above corresponds to the fattest possible ring.}. Figure \ref{BHfig} shows a $r-s$ plot of a static black hole with radius $R$, while figure \ref{BRfig} shows a $r-s$ plot of a black ring with $S^{D-3}$ radius (thickness) $R$ and $S^1$ radius (ring radius) $b$. To see why the condition $b>R$ is required when $b \neq 0$, rotate the $s$ vs $r$ curve through $2\pi$ around the $s$ axis ($r=0$) to get a $2D$ surface as the projection of the black ring in $(r,\theta,s)$ subspace. When $b<R$, i,e. the ring intersects the $s$ axis, this $2D$ surface has a conical singularity at $r=0$.\\

\subsection{Thin black ring in AdS background}\label{AdSring}

Now we consider the case $\Upsilon^2 = \frac{1}{l^2} \neq 0$ which is the AdS membrane with a specific nonzero rotation. The proof of nonexistence of black rings breaks down for such a rotation, so the existence of an AdS black ring is not ruled out. So this case is worth investigating for a black ring solution. We will attempt to solve \eqref{ChAdSdl} with this restriction.

\eqref{ChAdSdl} takes the form for $\Upsilon^2 = \frac{1}{l^2}$
\begin{equation}\label{ChAdSdlsp1}
\left(2h + (h')^2 + \frac{\left(2h-xh'\right)^2}{l^2}\right)\left(1+\frac{2h}{l^2}\right)
= \left(1+\frac{\left(2h-xh'\right)}{l^2}\right)^2\left(1+\frac{2h}{l^2}-\alpha^2\right)^2.
\end{equation}
This equation is very difficult to solve exactly. The quadratic ansatz for a ring,
\begin{equation}\label{BRansatz}
2h = p - q(x-c)^2,~~~~ q \neq 0,
\end{equation}
doesn't solve it either. So we use a thin ring approximation.

For a thin black ring, we explore the regime $c,l \gg 1$ with $c \sim l$. This translates into $L \gg R$ and $b = c\beta \gg R$, i.e. the $S^{D-3}$ scale is much smaller than the AdS radius and the $S^1$ radius, ergo, the `thin' ring. And we continue using the quadratic ansatz \eqref{BRansatz}. With this ansatz, \eqref{ChAdSdlsp1} becomes
\begin{equation}\label{ChAdSdlsp2}
\begin{split}
&\left(p-q(1-q)(x-c)^2+\frac{[(p+qc(x-c)]^2}{l^2}\right)\left(1+\frac{p-q(x-c)^2}{l^2}\right)\\
=& \left(1+\frac{p+qc(x-c)}{l^2}\right)^2\left(1-\alpha^2+\frac{p-q(x-c)^2}{l^2}\right)^2.
\end{split}
\end{equation}
Now let's see what simplification the thin ring regime has to offer. One easy way is to use a small parameter $\epsilon \ll 1$, declare $l$ and $c$ to be $\mathcal{O}(\epsilon^{-1})$ while $p$ and $q$ to be $\mathcal{O}(\epsilon^{0})$, and ultimately collect only $\mathcal{O}(\epsilon^{0})$ terms from \eqref{ChAdSdlsp2}, keeping in mind that $(x-c)$ also needs to be $\mathcal{O}(\epsilon^{0})$ since from \eqref{BRansatz}, $(x-c)^2 \leq \frac{p}{q}$. This yields
\begin{equation}\label{ChAdSdlsp3}
p - q\left(1-q-\frac{qc^2}{l^2}\right)(x-c)^2 = (1-\alpha^2)^2.
\end{equation}
Comparing the constant and the coefficient of $(x-c)^2$ on both sides,
\begin{equation}
p = (1-\alpha^2)^2,~~~~~~~~~ q = \left(1+\frac{c^2}{l^2}\right)^{-1}.
\end{equation}
Thus, we have obtained a charged thin AdS black ring solution rotating at $\omega = \frac{1}{L}$.

\subsection{Rotating uncharged black ring in flat background at the subleading order}\label{rotBRsubl}

Emparan and Reall \cite{Emparan:2001wk} first constructed the static black ring in 5 dimensions by extending the Weyl construction to $D > 4$. However they found that to keep the metric asymptotically flat, the static ring has to have a conical singularity in the near horizon metric. In \cite{Emparan:2001wk}, they showed that this singularity is removed when the ring is given an equilibrium rotation along $S^1$. This means in $5D$ the static ring doesn't exist, while rotating one does, which clearly contradicts the leading order result presented in \ref{BRlead}.

The exact black ring solution in $D > 5$ hasn't yet been found analytically. In \cite{Emparan:2007wm}, an approximate solution was constructed for any $D>5$ by compactifying a black string. The approximation is better for thin black ring (very large $S^1$ radius compared to the $S^{D-3}$ length scale, i.e. $\frac{R}{b}\ll 1$). The ring is thought to be sourced by a probe brane, and for equilibrium the ‘string tension’ has to be balanced by the centrifugal force since the self gravity is negligible in the thin ring approximation. This equilibrium angular velocity for a thin ring goes as $1 / \sqrt{D}$ as $D$ gets large. So the leading order membrane paradigm is unable to detect such a small rotation and the ring is perceived as static.

\cite{Tanabe:2015hda} obtained effective equations by assuming the angular velocity to be of the order $1 / \sqrt{D}$ a priori ,and found out the black ring solution with the desired rotation. We can't change the first order membrane equations to accommodate this scale of angular velocity, but we can go to the next order and see if we can recover such a rotation. The membrane equation in the flat case up to the first subleading order is given in \cite{Dandekar:2016fvw}:
\begin{equation}\label{2ndorderflat}
\begin{split}
&\color[rgb]{0.00,0.00,1.00}{\left[\frac{\nabla^2 u_\alpha}{ K}+u^\beta { K}_{\beta\alpha}-\frac{u^\beta { K}_{\beta \delta} { K}^\delta_\alpha}{ K}-\nabla_\alpha\ln K-u\cdot\nabla u_\alpha\right]P^\alpha_\gamma}\\
&
+ \Bigg[\frac{\nabla^2\nabla^2 u_\alpha}{{ K}^3}-\frac{(\nabla_\alpha{ K})(u\cdot\nabla{ K})}{{ K}^3}-\frac{(\nabla_\beta{ K})(\nabla^\beta u_\alpha)}{{ K}^2}-\frac{2{ K}^{\delta \sigma}\nabla_\delta\nabla_\sigma u_\alpha}{K^2} -\frac{\nabla_\alpha\nabla^2{ K}}{{ K}^3}
\\&+\frac{\nabla_\alpha({ K}_{\beta\delta} { K}^{\beta\delta} { K})}{K^3}+3\frac{(u\cdot { K}\cdot u)(u\cdot\nabla u_\alpha)}{{ K}}\\
&-3\frac{(u\cdot { K}\cdot u)(u^\beta { K}_{\beta\alpha})}{{ K}}
-6\frac{(u\cdot\nabla{ K})(u\cdot\nabla u_\alpha)}{{ K}^2}+6\frac{(u\cdot\nabla{ K})(u^\beta { K}_{\beta\alpha})}{{ K}^2}+3\frac{u\cdot\nabla u_\alpha}{D-3}\\
&-3\frac{u^\beta { K}_{\beta\alpha}}{D-3}\Bigg]P^\alpha_\gamma
=0.
\end{split}
\end{equation}
Let's solve this equation for a rotating configuration, perturbatively in rotation. Since $\partial_t$ and $\partial_\theta$ are both Killing vectors on a sphere, the blue terms can be rewritten as follows.\\
\begin{equation}\label{blu1}
\begin{split}
P^\alpha_\gamma\left[\frac{\nabla^2 u_\alpha}{ K}+u^\beta { K}_{\beta\alpha}-\frac{u^\beta { K}_{\beta \delta} { K}^\delta_\alpha}{ K}\right] &= \frac{1}K P^\alpha_\gamma\nabla^\mu\left(\nabla_\mu u_\alpha+\nabla_\alpha u_\mu\right),\\
&= \frac{1}K P^\alpha_\gamma\nabla^\mu\left(k_\alpha\nabla_\mu\gamma+k_\mu\nabla_\alpha\gamma\right),\\
&= \frac{1}K P^\alpha_\gamma\left[(\nabla_\mu k_\alpha) \nabla^\mu\gamma+k^\mu\nabla_\mu\nabla_\alpha\gamma\right],\\
&= \frac{2}K P^\alpha_\gamma k^\mu\nabla_\alpha\nabla_\mu\gamma,\\
&= \frac{2}{K}\left[k^\mu\nabla_\gamma\nabla_\mu\gamma+2u_\gamma(\nabla_\mu\ln\gamma)(\nabla^\mu\ln\gamma)\right].
\end{split}
\end{equation}
Here we have used
\begin{equation}
\begin{split}
k^\alpha k^\mu\nabla_\alpha\nabla_\mu\gamma &= (k\cdot\nabla)^2\gamma - (k\cdot\nabla k^\mu)\nabla_\mu\gamma,\\
&= \gamma^{-3}(\nabla_\mu\gamma)(\nabla^\mu\gamma).
\end{split}
\end{equation}
Also,
\begin{equation}
\begin{split}
P^\alpha_\gamma\left[\nabla_\alpha\ln K+u\cdot\nabla u_\alpha\right] &= P^\alpha_\gamma \nabla_\alpha\ln\left(\frac{K}\gamma\right),\\
&= \nabla_\gamma\ln\left(\frac{K}\gamma\right).
\end{split}
\end{equation}
So the blue terms become for a stationary configuration
\begin{equation}
\frac{2}K\left[k^\mu\nabla_\gamma\nabla_\mu\gamma+2u_\gamma(\nabla_\mu\ln\gamma)(\nabla^\mu\ln\gamma)\right]-\nabla_\gamma\ln\left(\frac{K}\gamma\right).
\end{equation}
Now let's consider the black ring solution
\begin{equation}
s^2-2g(r) = 0,
\end{equation}
where with the $\mathcal{O}(D^{-1})$ correction
\begin{equation}\label{shapesubl}
2g(r) = \beta^2 - (r-b)^2 + \frac{2\mathfrak{g}(r)}D.
\end{equation}
Let $k = \partial_t + \omega \partial_\theta$, and define
\begin{equation}
\omega = \sqrt{\frac{2}D}\tilde{\omega}.
\end{equation}
First, let's consider the $\mathcal{O}\left(D^{-\frac{1}2}\right)$. The black terms in \eqref{2ndorderflat} start at $\mathcal{O}\left(D^{-1}\right)$ and therefore don't contribute. While the contribution \eqref{blu1} to the blue terms depends on derivatives of $\gamma$ so it vanishes up to $\mathcal{O}\left(D^{-1}\right)$. So the effective equation for the singly rotating black object up to $\mathcal{O}\left(D^{-\frac{1}2}\right)$ is
\begin{equation}
\nabla_\mu\ln K = 0
\end{equation}
Since $\gamma$ is constant up to $\mathcal{O}\left(D^{-\frac{1}2}\right)$, so is $K$, Thus to this order the solution still has the $SO(D-2)$ symmetry.\\
Now let's consider the $\mathcal{O}\left(D^{-1}\right)$. Due to explicit suppression by $\frac{1}D$, in the black terms we have to use the leading order black ring solution. It can be easily seen that all of the black terms vanish for this configuration. And again, the contribution \eqref{blu1} to the blue terms vanishes up to $\mathcal{O}\left(D^{-1}\right)$. Thus at $\mathcal{O}\left(D^{-1}\right)$ the membrane equation again takes the form
\begin{equation}
\nabla_\mu\ln\left(\frac{K}\gamma\right) = 0,
\end{equation}
which has the solution $K = \frac{D\gamma}{\beta}$ i.e. to the order $D^0$
\begin{equation}
K = \frac{D}{\beta}\left(1+\frac{\tilde\omega^2r^2}D\right).
\end{equation}
It is solved in terms of the shape function in appendix \ref{subrot}. However the most striking result of that analysis is not the corrected shape function, but the condition on the angular velocity \eqref{rotcond}, which translates to
\begin{equation}\label{rotbeaut}
\omega = \frac{1}{\sqrt{D}~b}.
\end{equation}

\subsubsection*{Checking the angular velocity with the literature}

Now the angular velocity \eqref{rotbeaut} of the black ring has to be checked with that reported in \cite{Tanabe:2015hda}. In order to make this comparison, each solution in \cite{Tanabe:2015hda} should be identified with a solution in this section in a coordinate independent way, like equating the respective thermodynamic quantities. Unfortunately, \cite{Tanabe:2015hda} could not report the thermodynamic quantities in a closed form, because the integrations had to be done numerically, which they have done for $D = 14$\footnote{Since $n\equiv D-4$ is large, the integrations can be done using the saddle point approximation to find the mass and the horizon area, and  matching with the corresponding quantities from \ref{thermosec} gives the mapping $R = \frac{\tilde{R}}{\sqrt{\tilde{R}^2-1}}$ and $b = \sqrt{\tilde{R}^2-1}$, bringing us to the same conclusion as what follows from this section.}. So an attempt is made here to match the angular velocity via a physical argument.

In \cite{Tanabe:2015hda} the solutions are reported with a ``horizon scale'' fixed at unity. So we fix our $R = 1$ and identify the ring radius with that in \cite{Tanabe:2015hda}, i.e. $\tilde{R} = b$ \footnote{In \cite{Tanabe:2015hda} the ring radius is denoted by $R$, which clashes with our notation for the $S^{D-3}$ radius. To avoid confusion, the ring radius in \cite{Tanabe:2015hda} will be called $\tilde{R}$ here.}. Now we can compare temperatures of the matched solutions. The stationary membrane equation in \cite{Mandlik:2018wnw} for $\alpha = 0$ is
\begin{equation}
K = 4\pi T\gamma.
\end{equation}
For the angular velocity of $\mathcal{O}\left(1/\sqrt{D}\right)$, to the leading order $\gamma =1$. And the extrinsic curvature $K = \frac{D}{R} = D$. Thus the temperature of the membrane configuration is
\begin{equation}\label{Tempmem}
T = \frac{D}{4\pi},
\end{equation}
while the temperature reported in \cite{Tanabe:2015hda} is 
\begin{equation}\label{Temprep}
T = \frac{D}{4\pi}\frac{\sqrt{\tilde{R}^2-1}}{\tilde{R}}.
\end{equation}
These two solutions can be identified if there is a relative scaling of time, i.e. $t_M = \frac{\sqrt{\tilde{R}^2-1}}{\tilde{R}}t_T$, where $t_M$ and $t_T$ are the time coordinates in this paper and in \cite{Tanabe:2015hda} respectively. Since temperature is the inverse length of the thermal cycle when the time is Wick rotated, the scaling of temperature is equivalent of an inverse scaling of time. This means all the frequencies and angular velocities, which scale inversely with time, have to be multiplied by $\frac{\sqrt{\tilde{R}^2-1}}{\tilde{R}}$ when being converted from the membrane result to \cite{Tanabe:2015hda} result. (This would be useful later as well, when we compare the quasinormal mode frequencies with those of \cite{Chen:2017wpf} which uses the same methodology as that of \cite{Tanabe:2015hda}.) Thus the membrane angular velocity $\omega$ here gets converted to the ring equilibrium angular velocity $\Omega_{H}$ in \cite{Tanabe:2015hda} as
\begin{equation}\label{omegacomp}
\begin{split}
\Omega_{H} &= \frac{\sqrt{\tilde{R}^2-1}}{\tilde{R}}\omega,\\
&= \frac{\sqrt{\tilde{R}^2-1}}{\tilde{R}} \frac{1}{\sqrt{D}\tilde{R}},\\
&= \frac{1}{\sqrt{D}}\frac{\sqrt{\tilde{R}^2-1}}{\tilde{R}^2},
\end{split}
\end{equation}
which precisely matches the equilibrium angular velocity of \cite{Tanabe:2015hda}, the thin ring limit ($\tilde{R} \gg 1$) of which in turn matches the large $D$ limit of the arbitrary $D$ result of \cite{Emparan:2007wm}. These solutions are more general than the ones obtained in the large $D$ limit of \cite{Emparan:2007wm} since these rings need not be thin.

In \cite{Caldarelli:2008pz} the black ring with nonzero cosmological constant in $D > 4$ were constructed by the same principle as in \cite{Emparan:2007wm}: balancing the string tension at equilibrium. The equilibrium angular velocity of a thin AdS black ring was obtained as a $D$ dependent expression, and $D\to\infty$ limit of this expression yielded $\omega = \frac{1}{L}$ which precisely matches the result in \ref{AdSring}. If \eqref{ChAdSdlsp1} can be solved exactly, even in the uncharged ($\alpha = 0$) limit, it would give asymptotically AdS black rings which are not necessarily thin. However the equilibrium angular velocity will be the same for the fatter rings too, if they exist.

The static asymptotically de Sitter black ring in \cite{Caldarelli:2008pz} was obtained by balancing the string tension with cosmological constant. But this solution requires $b/L = 1/ \sqrt{D}$, which is out of the range of validity of the leading order large $D$ membrane paradigm with cosmological constant \cite{Bhattacharyya:2017hpj}, and therefore is invisible to the membrane paradigm, which the result of \ref{BRlead} suggests. It would be interesting to see if the subleading order correction to the membrane equations produces some black rings in nonzero cosmological constant case. If it doesn't, it would mean that a new regime of membrane paradigm with cosmological constant has to be worked out by scaling the cosmological constant with $D$ differently than the way it is done in \cite{Bhattacharyya:2017hpj, Bhattacharyya:2018szu, Kundu:2018dvx}. But if it does, that would result into novel asymptotically de Sitter non-thin black rings. 

\section{Comparison between thermodynamics of a black ring and a black hole}\label{thermosec}

Consider a thermal process that converts a black hole and a black ring into each other. As the large $D$ membrane does not radiate, the energy and charge should be conserved in these transformations. In other words, the energy and charge of both these configurations will be the same. So the more stable configuration will be the one that has the higher number of microstates, i.e. larger entropy for given energy and charge.\\
For a static configuration of a membrane in flat background, $\gamma = 1$, so the stationary membrane equations \eqref{stMeq} read
\begin{equation}
Q=\alpha,~~~ K  = \frac{D}{\beta(1-\alpha^2)},
\end{equation}
where $\alpha = 2\sqrt{2\pi}\mu$ and $\beta = \frac{D}{4\pi T}$ are constants, thus $Q$ and $ K $ are constant for a given static configuration. 

The energy of a static configuration is given by
\footnote{This note uses the units in which $16\pi G = 1$ while the previous work in \cite{Mandlik:2018wnw} uses $G=1$. See appendix \ref{units} for this conversion.}
\begin{equation}
\begin{split}
M &= \int dA~T_{00},\\
&= \int dA~ K (1+Q^2),\\
&=  K (1+\alpha^2)A,
\end{split}
\end{equation}
and the total charge of such a configuration is
\begin{equation}
\begin{split}
q &= \int dA~J_0,\\
&= 4\sqrt{2\pi}\int dA~ K Q,\\
&= 4\sqrt{2\pi}\alpha K A.
\end{split}
\end{equation}
$T_{00}$ and $J_0$ are obtained from \eqref{Thermoq}.

Now consider a charged black hole and a charged black ring in flat background. Let $T$ and $\bar{T}$ be their respective temperatures and $Q$ and $\bar{Q}$ be their respective charge parameters. Let 
$$\beta = \frac{D}{4\pi T}~~{\rm and}~~\bar\beta = \frac{D}{4\pi\bar{T}}.$$
and
$$\alpha = \frac{Q}{2\sqrt{2\pi}}~~{\rm and}~~\bar\alpha = \frac{\bar{Q}}{2\sqrt{2\pi}}.$$

Let $A$ and $\bar{A}$ be their respective areas, and $K$ and $\bar{K}$ be their respective scalar curvatures. Then the requirement that the charge of both the configurations is the same gives
\begin{equation}
\alpha K A = \bar{\alpha}\bar{ K }\bar{A},
\end{equation}
and the requirement that the energy of both the configurations is the same gives
\begin{equation}
 K A(1-\alpha^2) = \bar{ K }\bar{A}(1-\bar\alpha^2).
\end{equation}
These two equations give $\alpha = \bar\alpha$ and
\begin{equation}\label{Eqen}
 K A = \bar{ K }\bar{A}.
\end{equation}

In $(t,r,\theta,s,\{\chi^a\})$ coordinates the black hole is given by
\begin{equation}
s^2 + r^2 = R^2,
\end{equation}
and the black ring is given by
\begin{equation}
s^2 + (r-b)^2 = \bar{R}^2,
\end{equation}
where $R = \beta(1-\alpha^2)$ and $\bar{R} = \bar\beta(1-\bar\alpha^2)$.\\
The scalar extrinsic curvatures are obtained by
\begin{equation}
 K  = \frac{Dn_s}{s}.
\end{equation}
Thus,
\begin{equation}\label{KR}
 K  = \frac{D}{R},~~~\bar{ K } = \frac{D}{\bar{R}}~.
\end{equation}

The respective areas of black hole and black ring can be easily calculated by going to the $(t,\sigma,\phi, \theta,\{\chi^a\})$ coordinates (defined in section \ref{lightQNM} \footnote{Although this coordinate system is defined for a black ring, it can be used for a spherical black hole by putting $b=0$ and changing the range of $\phi$ to $[0,\frac{\pi}2]$}) in the respective cases, and noting that the membranes are located at $\sigma = 1$,
\begin{equation}\label{Area}
\begin{split}
A &= 2\pi\Omega_{D-4}R^{D-2}\int_0^{\frac{\pi}2}(\cos\phi)^{D-4}\sin\phi d\phi = \frac{2\pi\Omega_{D-4}}D R^{D-2},\\
\bar{A} &=  2\pi\Omega_{D-4}\bar{R}^{D-3}\int_{-\frac{\pi}2}^{\frac{\pi}2}(\cos\phi)^{D-4}(\bar{R}\sin\phi+b) d\phi = 2\pi b~\Omega_{D-4}\sqrt{\frac{2\pi}D} \bar{R}^{D-3}, 
\end{split}
\end{equation}

Using \eqref{KR} and \eqref{Area}, the condition \eqref{Eqen} now reads
\begin{equation}
R^{D-3} = \bar{R}^{D-4}b\sqrt{2\pi D}~.
\end{equation}
Substituting back into \eqref{Area}
\begin{equation}
\frac{\bar{A}}{A} = \left(\frac{\bar{R}}{b\sqrt{2\pi D}}\right)^{\frac{1}{D-3}},
\end{equation}
and the respective entropies are given by
\begin{equation}
S = 4\pi A,~~~~~~~~~~~~~~~~~\bar{S} = 4\pi\bar{A}.\\
\end{equation}
Thus the two entropies can be compared:
\begin{equation}\label{entratio}
\frac{\bar{S}}{S} = \frac{\bar{A}}{A} = \left(\frac{\bar{R}}{b\sqrt{2\pi D}}\right)^{\frac{1}{D-3}}=1-\frac{\ln D}{2D}+\mathcal{O}\left(D^{-1}\right).
\end{equation}
For a black ring with $b$ and $R$ of order $D^0$, the above ratio is smaller than $1$, up to $\mathcal{O}\left(\frac{\ln D}{D}\right)$\footnote{At first, it may seem that the subleading order corrections in $D$ should change this argument. But the correction to \eqref{entratio} due to the subleading order turns out to be $\mathcal{O}\left(D^{-1}\right)$ which is negligible compared to $\mathcal{O}\left(\frac{\ln D}{D}\right)$ at large $D$.}. This means the black hole has more number of configurations than any black ring configuration of given energy and total charge, and hence is thermodynamically more stable.

Note that in the strict $D \to \infty$ limit the ratio of entropies goes to $1$. However for a large but finite $D$, the black hole is more stable.

In a grand canonical Ensemble however, all the configurations have zero free energy at the leading order in $D$ and thus have equal thermodynamics stability. See the appendix \ref{grandcan}.

\subsection{Towards the black ring phase structure}\label{thermocomp}

In \cite{Emparan:2006mm}, Emparan and Reall obtained the phase structure for $5D$ black ring and rotating Myers-Perry (MP) black hole. The dimensionless horizon area $a_H$ was plotted against dimensionless angular momentum $j$ for both cases (fig.(2) in \cite{Emparan:2006mm}), and they found out that a ‘cusp’ divides the $a_H$ vs $j$ curve of the black ring into two branches: “thin” and “fat” black ring. The cusp lies at the minimum $j$ black ring can have, which is less than the maximum $j$ a MP black hole can have. Thus there is a range of $j$ where a MP black hole, a thin and a fat ring coexist. In this range, MP black hole has the highest entropy ($a_H$) while the fat ring has the lowest. For $j$ larger than the maximum for MP black hole, only the thin black ring exists (thin black ring branch has a large $j$ tail).

For $D > 5$ the MP black hole can have arbitrarily high $j$. So the black ring and MP black hole coexist at arbitrarily high $j$ as well. In \cite{Emparan:2007wm} the $a_H$ vs $j$ curves were compared, and again near the cusp (min. $j$ for black ring), MP black hole has more entropy. (The cusp was conjectured since the solution was only obtained for the thin ring). So since the angular momentum decreases as $1 / \sqrt{D}$ with $D$, at large $D$ the cusp should lie close to $j=0$ and below the curve for MP black hole, and hence the black hole should be thermodynamically favoured over the ring, which agrees with the result in this section.

The analysis of this section could be replicated for comparing the thin AdS black ring of \ref{AdSring} with a singly rotating AdS black hole \cite{Mandlik:2018wnw}. It is left for future work.

\section{Light Quasinormal modes about a static black ring in flat background}\label{lightQNM}

Now let's study the dynamical stability of the black ring solution at the leading order, namely \eqref{brsol}, by solving linearized membrane equation in \cite{Bhattacharyya:2015fdk} for small fluctuations about this solution. In the $(t,r,\theta,s,\{\chi^a\})$ coordinate system the metric is given by

\begin{equation}
-d\tau^2 = -dt^2 + dr^2 + r^2d\theta^2 + ds^2 + s^2d\Omega_{D-4}^2,
\end{equation}
where $-\pi \leq \theta < \pi$, with $\theta = \pm \pi$ identified with periodic boundary conditions, and $r\geq 0$, $s\geq 0$. Let
\begin{equation}
s = R\sigma \cos\phi,~~~~ r = R\sigma \sin\phi + b,
\end{equation}
with $-\frac{\pi}2 \leq \phi \leq \frac{\pi}2$ and $\sigma\geq 0$. In this coordinate system the unperturbed membrane is located at $\sigma = 1$ and the metric becomes
\begin{equation}\label{bgdmet}
-d\tau^2 = -dt^2 + R^2\bigg(d\sigma^2 + \sigma^2d\phi^2+ \sigma^2C^2d\Omega_{D-4}^2\bigg)+(R\sigma S+b)^2d\theta^2,
\end{equation}
where $S \equiv \sin\phi$ and $C \equiv \cos\phi$.

Let $X^M \equiv (t,\sigma,\phi,\theta,\{\chi^a\})$ be the coordinates of the background flat spacetime where $\chi^a$ are the angles on the $S^{D-4}$. $y^{\mu} \equiv (t,\phi,\theta,\{\chi^a\})$ are chosen to be the coordinates on the membrane. On the unperturbed membrane ($\sigma = 1$), such a point corresponds to a point $(t,1,\phi,\theta,\{\chi^a\})$ of the spacetime.

Now let's perturb the membrane by shifting the $\sigma$ at each point $y$ on it from $1$ to $1+\epsilon~\delta\rho(y)$. In other words, the membrane is now given to the order $\epsilon^1$ by
\begin{equation}
\sigma - \epsilon~\delta\rho(y) = 1.
\end{equation}
The outward pointing normal on this membrane
\begin{equation}
\begin{split}
n_{\sigma} &= R,\\
n_{\mu} &= -\epsilon R\partial_\mu\delta\rho.
\end{split}
\end{equation}
The bar indicates that the quantity is evaluated at $\sigma = 1$. If any $\sigma$ derivatives are acting on it, they act before setting $\sigma = 1$.

The pullback factors from the spacetime to the membrane are
\begin{equation}
e^{\sigma}_{\mu} = \epsilon\partial_{\mu}\delta\rho,~~~~~e^\mu_\nu = \delta^\mu_\nu.
\end{equation}

The nonzero Christoffel symbols in spacetime, $\Sigma^!_{BC}$ are given in \eqref{stchr}, which are used to calculate the components of the extrinsic curvature of the membrane, $K_{\mu\nu} = e^M_\mu e^N_\nu \nabla_M n_N$:
\begin{equation}
\begin{split}
K_{t\mu} &= -\epsilon R\partial_t\partial_\mu\delta\rho,\\
K_{\phi\phi} &=  R\left(1+\epsilon\delta\rho\right)-\epsilon R\partial_\phi^2\delta\rho,\\
K_{\phi\theta} &= -\epsilon R\left(\partial_\phi\partial_\theta\delta\rho - \frac{RC}{RS + b}\partial_\theta\delta\rho\right),\\
K_{\phi a} &= -\epsilon R\left(\partial_\phi\partial_a\delta\rho + \frac{S}C\partial_a\delta\rho\right),\\
K_{\theta\theta} &= S(RS(1+\epsilon\delta\rho) +b) - \epsilon R \partial_\theta^2\delta\rho - \epsilon C(RS+b)\partial_\phi\delta\rho,\\
K_{\theta a} &= -\epsilon R \partial_\theta\partial_a \delta\rho,\\
K_{ab} &= R(1+\epsilon\delta\rho)C^2\hat{g}_{ab} -\epsilon R\hat\nabla_a\hat\nabla_b \delta\rho + \epsilon RSC \partial_\phi \delta\rho\hat{g}_{ab},\\
K &= \frac{D}{R}(1-\epsilon\delta\rho)-\frac{\epsilon}{RC^2}\hat\nabla^2\delta\rho + D\frac{\epsilon S}{RC} \partial_\phi \delta\rho + {\cal O}\left(D^0\right).
\end{split}
\end{equation}
The covariant derivatives and Christoffel symbols with a hat on top indicate that they are taken on a unite $S^{D-3}$.

The induced metric on the membrane $g_{\mu\nu} \equiv e^M_\mu e^N_\nu G_{MN}$ is
\begin{equation}\label{indmet}
\begin{split}
-d\tau_{ind}^2 =& -dt^2 + R^2\bigg((1+2\epsilon\delta\rho)d\phi^2+ (1+2\epsilon\delta\rho)C^2d\Omega_{D-4}^2\bigg)\\
&+\bigg(R(1+\epsilon\delta\rho)S+b\bigg)^2d\theta^2.
\end{split}
\end{equation}

The Christoffel symbols for \eqref{indmet}, $\Gamma^\mu_{\alpha\beta}$, are given in \eqref{memchr0} and \eqref{memchr1}. The unperturbed induced metric and nonzero Christoffel symbols are denoted by a bar on top and obtained by putting $\delta\rho = 0$ (or effectively $\epsilon = 0$) in \eqref{indmet} and \eqref{memchr0} respectively.

The unperturbed velocity is given by
$$\bar{u}(y) = \partial_t.$$

Now we perturb the velocity
$$u^\mu(y) = \bar{u}^\mu(y) + \epsilon\delta u^\mu(y),~~~~\epsilon \ll 1.$$
$n\cdot u = u^\sigma = 0$ by above definition. $u\cdot u = -1$ yields $\delta u^t = 0$. And $\tilde\nabla\cdot u = 0$ (the tilde denotes that the divergence is taken on the membrane) gives
\begin{equation}\label{divucond}
\frac{\hat\nabla_a \delta u^a}{D}-\frac{S}{C}\delta u^\phi +\partial_t\delta\rho=0.
\end{equation}
Which means $u^\phi$ fluctuations are sourced by $u^a$ fluctuations and by shape fluctuations. These two sources can be turned on independently of each other.

The charge fluctuations are simply given by
\begin{equation}
Q(y) = \alpha + \epsilon\delta Q(y),
\end{equation}
where $\alpha$ is constant over the membrane. 

\subsection{Linearised membrane equation}\label{LQNM}

\begin{equation}\label{Veq}
{\cal E}^\mu \equiv \left(\frac{\tilde{\nabla}^2 u^\nu}{K}-(1-Q^2)\partial^\nu \ln K + u\cdot K^\nu -(1+Q^2)u\cdot\tilde{\nabla}u^\nu\right)p_\nu^\mu = 0,
\end{equation}
and
\begin{equation}\label{Seq}
{\cal E}_S\equiv\frac{\tilde\nabla^2\ln Q}{K} -u\cdot\partial \ln(QK) +u\cdot K\cdot u = 0.
\end{equation}
It can be easily shown that
\begin{equation}
\begin{split}
p^\mu_\nu\tilde\nabla^2 u^\nu &= \tilde{\nabla}^2 u^\mu + \mathcal{O}(D^0)\\
&= \frac{\hat{g}^{ab}}{C^2}\left(\Gamma^\mu_{ac}\Gamma^c_{b\nu}u^\nu-\Gamma^\nu_{ab}(\partial_\nu u^\mu +\Gamma^\mu_{\nu\alpha}u^\alpha)\right) + \mathcal{O}(D^0)
\end{split}
\end{equation}

Now let's check that the unperturbed configuration solves these equations. Setting all fluctuations to zero, \eqref{Seq} is satisfied identically as $Q$ and $K$ are constants while $K_{tt} = 0$. Also, in \eqref{Veq} each of the terms inside the bracket vanishes individually. This also tells us that the variation of \eqref{Veq} with charge also identically vanishes.

Upon turning on the velocity, shape and charge fluctuations the membrane equations take form
\begin{equation}
\mathcal{E}_\mu = \epsilon\delta\mathcal{E}_\mu,~~~~~~~~~~\mathcal{E}_S = \epsilon\delta\mathcal{E}_S~,
\end{equation}
where
\begin{equation}
\begin{split}
\delta\mathcal{E}_t =& 0,
\\
\delta\mathcal{E}^\theta =& \frac{1}{R}\left(\frac{\hat\nabla^2}{DC^2}-\frac{S}{C}\partial_\phi-\frac{RS}{RS+b}-(1+\alpha^2)R\partial_t+\frac{RS}{RS+b}\right)\delta u^\theta\\
&+\frac{1}{(RS+b)^2}\partial_\theta\left((1-\alpha^2)\left(\frac{\hat\nabla^2}{DC^2}-\frac{S}{C}\partial_\phi+1\right)-R\partial_t\right)\delta\rho,
\\
\delta\mathcal{E}^\phi =& \frac{1}{R}\left(\frac{\hat\nabla^2}{DC^2}-\frac{S}{C}\partial_\phi-\frac{S^2}{C^2}-(1+\alpha^2)R\partial_t+1\right)\delta u^\phi+\frac{2S}{RC}\frac{\hat\nabla_a\delta u^a}{D}\\
&+\frac{1}{R^2}\partial_\phi\left((1-\alpha^2)\left(\frac{\hat\nabla^2}{DC^2}-\frac{S}{C}\partial_\phi+1\right)-R\partial_t\right)\delta\rho,
\\
\delta\mathcal{E}^a =& \frac{1}{R}\left(\frac{\hat\nabla^2}{DC^2}-\frac{S}{C}\partial_\phi+\frac{S^2}{C^2}-(1+\alpha^2)R\partial_t+1\right)\delta u^a\\
&+\frac{1}{R^2C^2}\hat\nabla_a\left((1-\alpha^2)\left(\frac{\hat\nabla^2}{DC^2}-\frac{S}{C}\partial_\phi+1\right)-R\partial_t\right)\delta\rho,
\end{split}
\end{equation}
which upon simplification and use of \eqref{divucond} gives
\begin{equation}\label{lineqV}
\begin{split}
\delta\mathcal{E}_t =& 0,
\\
\delta\mathcal{E}^\theta =& \frac{1}{R}\left(\frac{\hat\nabla^2}{DC^2}-\frac{S}{C}\partial_\phi-(1+\alpha^2)R\partial_t\right)\delta u^\theta\\
&+\frac{1}{(RS+b)^2}\partial_\theta\left((1-\alpha^2)\left(\frac{\hat\nabla^2}{DC^2}-\frac{S}{C}\partial_\phi+1\right)-R\partial_t\right)\delta\rho,
\\
\delta\mathcal{E}^\phi =& \frac{1}{R}\left(\frac{\hat\nabla^2}{DC^2}-\frac{S}{C}\partial_\phi-(1+\alpha^2)R\partial_t+\frac{1}{C^2}\right)\delta u^\phi-\frac{2S}{RC}\partial_t\delta\rho\\
&+\frac{1}{R^2}\partial_\phi\left((1-\alpha^2)\left(\frac{\hat\nabla^2}{DC^2}-\frac{S}{C}\partial_\phi+1\right)-R\partial_t\right)\delta\rho,
\\
\delta\mathcal{E}^a =& \frac{1}{R}\left(\frac{\hat\nabla^2}{DC^2}-\frac{S}{C}\partial_\phi-(1+\alpha^2)R\partial_t+\frac{1}{C^2}\right)\delta u^a\\
&+\frac{1}{R^2C^2}\hat{g}^{ab}\partial_b\left((1-\alpha^2)\left(\frac{\hat\nabla^2}{DC^2}-\frac{S}{C}\partial_\phi+1\right)-R\partial_t\right)\delta\rho,
\end{split}
\end{equation}
and
\begin{equation}\label{lineqS}
\delta\mathcal{E}_S \equiv \frac{1}R\left(\frac{\hat{\nabla}^2}{DC^2}-\frac{S}{C}\partial_\phi\right)\frac{\delta Q}{\alpha}-\partial_t\left(\frac{\delta Q}{\alpha}-\left(\frac{\hat{\nabla}^2}{DC^2}-\frac{S}{C}\partial_\phi+1\right)\delta\rho\right)-R\partial_t^2\delta\rho = 0.
\end{equation}

\subsubsection{QNM for $\delta\rho$ fluctuation}
Evaluation of $\frac{\tilde\nabla\cdot\mathcal{E}}{D} = 0$, i.e. $\frac{\hat\nabla_a\mathcal{E}^a}{D}-\frac{S}{C}\mathcal{E}^\phi=0$ gives
\begin{equation}\label{divdeltu}
\begin{split}
&\frac{1}{R}\left(\frac{\hat\nabla^2}{DC^2}-\frac{S}{C}\partial_\phi-(1+\alpha^2)R\partial_t+\frac{1}{C^2}\right)\frac{\hat\nabla_a\delta u^a}{D}+\frac{1}{RC^2}\frac{\hat\nabla_a\delta u^a}{D}\\
&-\frac{S}{RC}\left(\frac{\hat\nabla^2}{DC^2}-\frac{S}{C}\partial_\phi-(1+\alpha^2)R\partial_t+\frac{1}{C^2}\right)\delta u^\phi+\frac{2S^2}{RC^2}\partial_t\delta\rho\\
&+\frac{1}{R^2}\left(\frac{\hat\nabla^2}{DC^2}-\frac{S}{C}\partial_\phi\right)\left((1-\alpha^2)\left(\frac{\hat\nabla^2}{DC^2}-\frac{S}{C}\partial_\phi+1\right)\right)\delta\rho~~=~~0.
\end{split}
\end{equation}
Again using \eqref{divucond} and simplifying we get
\begin{equation}\label{rhofluc}
\left((1-\alpha^2)\mathcal{D}(\mathcal{D}+1) - 2R(\mathcal{D}+1)\partial_t + R^2(1+\alpha^2)\partial_t^2\right)\delta\rho = 0,
\end{equation}
where
\begin{equation}\label{fancyD}
\mathcal{D} \equiv \frac{\hat\nabla^2}{DC^2}-\frac{S}{C}\partial_\phi.
\end{equation}
is the laplacian on $S^{D-3}$ (see \ref{SD-3}).
This equation is independent of charge and velocity fluctuations, which demonstrates that shape can be perturbed independently. However, the charge and the velocity fluctuations are also forced by the shape fluctuations according to \eqref{lineqS} and \eqref{lineqV} respectively.\\
Let us expand $\delta\rho$ in a basis of regular scalars on the black ring
\begin{equation}\label{rhodec}
\delta\rho = \sum_{L=0}^{\infty} f^\rho_L(\theta)Y^{(L)}e^{-i\omega_L^{(\rho)} t},~~~L \geq 0.
\end{equation}
\footnote{Apologies to the reader for recycling the label $L$ for the harmonic number, which was earlier used for AdS radius.} Where $Y^{(L)}$ are the scalar spherical harmonics on $S^{D-3}$, which should be the eigenfunctions of \eqref{rhofluc} because $\mathcal{D}$ is proportional to the laplacian on $S^{D-3}$. Since the equation \eqref{rhofluc} doesn't depend on $\theta$ or $\partial_\theta$, every $f_L(\theta)$ can be any arbitrary function that respects the periodicity of $\theta$. Also, $D\mathcal{D}$ is the laplacian acting on a scalar on $S^{D-3}$, thus
$$\mathcal{D}Y^{(L)} = -LY^{(L)} + \mathcal{O}(D^0)$$.
so using \eqref{rhodec} in \eqref{rhofluc} gives
\begin{equation}
R^2(1+\alpha^2){\omega_L^{(\rho)}}^2+2iR{\omega_L^{(\rho)}}(L-1)-(1-\alpha^2)L(L-1) = 0,
\end{equation}
which can be solved for $\omega_L^{(\rho)}$ to get the quasinormal frequencies for the shape fluctuations:
\begin{equation}
\omega_{L,\pm}^{(\rho)} = \frac{-i(L-1)\pm\sqrt{(1-\alpha^4L)(L-1)}}{R(1+\alpha^2)},~~~L \geq 0.
\end{equation}

\subsection{QNM for $\delta u$ fluctuation}
As seen from \eqref{lineqV}, the velocity fluctuations are sourced by the shape fluctuations. But the kernel of the differential operator acting on the velocity fluctuations is obtained simply by setting $\delta\rho = 0$ in \eqref{lineqV}, which is spanned by the quasinormal modes of the velocity fluctuations. Thus, these modes satisfy
\begin{equation}\label{velQNM}
\begin{split}
& \left(\frac{\hat\nabla^2}{DC^2}-\frac{S}{C}\partial_\phi-(1+\alpha^2)R\partial_t\right)\delta u^\theta = 0,
\\
& \left(\frac{\hat\nabla^2}{DC^2}-\frac{S}{C}\partial_\phi-(1+\alpha^2)R\partial_t+\frac{1}{C^2}\right)\delta u^\phi=0,
\\
& \left(\frac{\hat\nabla^2}{DC^2}-\frac{S}{C}\partial_\phi-(1+\alpha^2)R\partial_t+\frac{1}{C^2}\right)\delta u^a=0.
\end{split}
\end{equation}
When written in terms of $S^1 \times S^{D-3}$ split form of the velocity fluctuations $(\delta u^\theta,\delta u^{\Xi})$ as given in \ref{SD-3}, \eqref{velQNM} take the form
\begin{equation}\label{velQNMSD-3}
\begin{split}
\left(\mathcal{D}-(1+\alpha^2)R\partial_t\right)\delta u^\theta &= 0,\\
\left(\mathcal{D}_V+1 - (1+\alpha^2)R\partial_t\right)\delta u^\Xi &= 0.
\end{split}
\end{equation}
where $D\mathcal{D}_V$ is the laplacian acting on vectors on $S^{D-3}$, see \ref{SD-3}.\\
We expand the velocity fluctuations in the basis
\begin{equation}
\begin{split}
\delta u^{\theta} &= \sum_{L=0}^{\infty}f^\theta_L(\theta)Y^{(L)}e^{-i\omega_L^{(\theta)} t},\\
\delta u^{\Xi} &= \sum_{L=1}^{\infty}f^{V}_L(\theta)V_{(L)}^{\Xi}e^{-i\omega_L^{(V)} t},
\end{split}
\end{equation}
where $V_{(L)}$ are the vector spherical harmonics on $S^{D-3}$ (again, as justified in \ref{SD-3}). Then \eqref{velQNMSD-3} gives the quasinormal frequencies for the velocity fluctuations:
\begin{equation}\label{velfreq}
\begin{split}
\omega_L^{(\theta)} &= \frac{-iL}{R(1+\alpha^2)},~~~L \geq 0,\\
\omega_L^{(V)} &= \frac{-i(L-1)}{R(1+\alpha^2)},~~~L \geq 1.
\end{split}
\end{equation}

\subsection{QNM for $\delta Q$ fluctuation}
Like the velocity fluctuations, the charge fluctuations are sourced by shape fluctuations too. Here also, the quasinormal modes for the charge fluctuations can be obtained simply by setting $\delta\rho = 0$ in \eqref{lineqS}, which gives the equation for these modes
\begin{equation}\label{chQNM}
\left(\frac{\hat{\nabla}^2}{DC^2}-\frac{S}{C}\partial_\phi-R\partial_t\right)\delta Q =0.
\end{equation}
In other words,
\begin{equation}\label{chQNMSD-3}
\left(\mathcal{D}-R\partial_t\right)\delta Q = 0.
\end{equation}
Like the shape fluctuations, we expand the charge fluctuations as
\begin{equation}
\delta Q = \sum_{L=0}^{\infty}f^\theta_L(\theta)Y^{(L)}e^{-i\omega_L^{(Q)} t},~~~L \geq 0.
\end{equation}
When substituted in \eqref{chQNMSD-3}, this yields the quasinormal frequencies for charge fluctuations
\begin{equation}\label{chfreq}
\omega_L^{(Q)} = \frac{-iL}{R},~~~L \geq 0.
\end{equation}

\subsection{Stability of black ring membrane and interpretation of zero modes}\label{stabstab}

All of these quasinormal frequencies have a negative semidefinite imaginary part. So none of the modes listed in this section has an exponential growth with time. Which means the black ring \eqref{brsol} has no unstable light quasinormal modes.\footnote{The $L=0$ modes for the shape fluctuation having the frequency $\omega_{0,+}^{(\rho)}$ look like blowing up as each of them a positive imaginary part, but they can be deemed unphysical exactly in the same manner as the similar mode in black hole membrane was argued to be unphysical in \cite{Bhattacharyya:2015fdk}.}

This, however, doesn't mean that the black ring is stable. The linear analysis of the first order membrane equations as done above concerns with only the light quasinormal modes, which have frequencies of the order $D^0$. The analysis in \cite{Chen:2017wpf} (and earlier in \cite{Tanabe:2015hda} for the uncharged case) obtained unstable modes with frequencies of order $\sqrt{D}$ (called ``non-axisymmetric''\footnote{The nomenclature ``axisymmetric'' and ``non-axisymmetric'' modes in \cite{Tanabe:2015hda} and \cite{Chen:2017wpf} doesn't reflect whether the modes are $\theta$ dependent or not, it's due to a peculiar scaling of the angular coordinate on $S^1$ used there.} modes there), which they claimed to correspond to the Gregory-Laflamme type instability \cite{Gregory:1993vy, Gregory:1994bj}. \cite{Dandekar:2016jrp} showed a way to obtain such Gregory-Laflamme modes for the black branes by considering the order $1/ \sqrt{D}$ wavelength fluctuations. It would be interesting to see if that analysis can be adapted to black rings to get such modes. This can be done at the leading order itself.  

The frequencies of “axisymmetric” ($m = 0$) modes of \cite{Chen:2017wpf} fall in the range of light quasinormal modes. The gravitational quasinormal modes in \cite{Chen:2017wpf} are
\begin{equation}
\omega^{m=0}_\pm = \frac{\sqrt{\tilde{R}^2-1}}{\tilde{R}}\left(\frac{-(l-1)\pm\sqrt{(l-1)(1-\alpha^4l)}}{(1+\alpha^2)}\right)~,
\end{equation}
and the charge perturbation quasinormal modes are
\begin{equation}
\omega_c = -il~\frac{\sqrt{\tilde{R}^2-1}}{\tilde{R}},
\end{equation}
both of which match with the results in this section for $R=1$ with the ``time scaling'' as argued in \ref{rotBRsubl} (and identifying $l = L$).

Since \cite{Tanabe:2015hda} and \cite{Chen:2017wpf} deal entirely with the scalar fluctuations which play a role in the GL-like instabilities, they don't have the spectrum of modes corresponding to “velocity perturbations” that is obtained here. So these modes is a novel result. Also, the Gregory-Laflamme instability analysis hasn't been done for the five dimensional fat black rings. (they however are shown to be unstable to radial fluctuations, \cite{Elvang:2006dd}), whereas according to \cite{Tanabe:2015hda} and \cite{Chen:2017wpf} all black rings at large $D$ have this instability, irrespective of their thickness. Can membrane paradigm see the Gregory-Laflamme instability of the fat rings in large $D$?

The most striking feature of the light quasinormal modes presented here is the presence of an infinite number of zero modes. The quasinormal frequencies $\omega_{0,-}^{(\rho)}$, $\omega_{1}^{(\rho)}$ $\omega_{0}^{(Q)}$, $\omega_{0}^{(\theta)}$ and $\omega_{1}^{(V)}$ are all zero. This means the black ring slightly perturbed with the modes corresponding to these frequencies are also stationary solutions of the membrane equations. The black hole membrane of \cite{Bhattacharyya:2015fdk} also had such zero modes, but they were trivial in the sense that they corresponded to small uniform changes in size, charge, small displacement and boost of the black hole, which were trivially the solutions. But here, these zero modes are only uniform in the $S^{D-3}$ directions, not in $\theta$, because of the arbitrary functions of $\theta$ multiplying these modes. So such deformations can give rise to non-axisymmetric black rings. Some specific non-axisymmetric black rings are shown to be the end states of Gregory-Laflamme instability of black rings in $5D$ \cite{Horowitz:2001cz, Figueras:2015hkb}. However, large enough deformations may not be stationary due to nonlinear effects, and their dynamics might reveal the dynamics of nonuniform black rings towards the end states as established numerically in \cite{Chen:2018vbv}.
\section{Conclusion}\label{theend}

Let's first summarize the main results of this paper:
\begin{itemize}
 \item A simple analysis of the first order stationary membrane equations of \cite{Mandlik:2018wnw}, specialized to axisymmetric configurations reveals, even without finding explicit solutions, that these equations admit a black ring topology ($S^1 \times S^{D-3}$) of the membrane in flat space only if it is static, both in the charged and in the uncharged case. 
 \item However, the subleading order correction to the equations produces an asymptotically flat rotating uncharged black ring with the angular velocity reported in \cite{Emparan:2007wm} and \cite{Tanabe:2015hda}.
 \item The thin AdS black ring of \cite{Caldarelli:2008pz} is replicated, and it is shown that any AdS black ring can exist only at a particular angular velocity, $\omega = \frac{1}{L}$ irrespective of the dimensions of and the charge on the black ring.
 \item The de Sitter black ring lies outside the domain of applicability of first order membrane paradigm.
 \item At arbitrarily large but finite $D$, the static asymptotically flat black ring is thermodynamically less favoured compared the the flat Reissner-Nordstrom black hole in the microcanonical ensemble. When $D \to \infty$, however, both configurations have equal entropy. In Grand canonical ensemble, both configurations have zero free energy.
 \item The light quasinormal mode frequencies of the charged black ring membrane \eqref{brsol} were found and shown to match with the ``axisymmetric'' sector mode frequencies of \cite{Chen:2017wpf}. None of these modes is unstable.
 \item There exist infinitely many axisymmetry breaking zero modes which can produce nonuniformity in the black ring. These may lead to nonuniform black ring solutions which are the end-points of GL like instabilities, as demonstrated in \cite{Chen:2018vbv}.
\end{itemize}

These results open up several directions of investigation which are already mentioned in this text. The subleading order corrections to the membrane equations can produce the rotating charged black ring, and the black rings in de Sitter. An analysis of membrane equations in $1/ \sqrt{D}$ perturbation can help studying Gregory-Laflamme instability of the black rings.

In \cite{Tanabe:2015hda} and \cite{Chen:2017wpf} the black rings were constructed in the ``ring coordinates''. These rings were compared to those in this work by employing physical arguments. But it would be good to construct the membrane in this coordinate system so that the comparison is clearer. This would especially be very useful in the AdS case. In the axisymmetric coordinate system used in the membrane paradigm, the temperature doesn't depend of $b$ which parametrizes the ring radius, whereas in the ring coordinates it does. The ring coordinates are dependent on the ring radius unlike the axisymmetric coordinate system, and that might be bringing out this ring radius dependence. In this paper, the equilibrium angular velocity too is shown to be independent of $b$ in the AdS case. It would be interesting to see how this result looks like in the coordinate system specifically designed for black rings.

An important long term goal would be to obtain the significant part of the black hole phase diagram at large $D$. A simpler immediate goal would be to study the singly rotating sector of this diagram, as done in \cite{Emparan:2007wm} in $5$ dimensions. The singly rotating black hole solutions (topology $S^{D-2}$) were already obtained in \cite{Mandlik:2018wnw}. With the rotating black rings we can compare the $a_H$ vs $j$ curves of the two topologies, like in fig.(2) of \cite{Emparan:2007wm}.

A significant portion of the phase space of black objects is covered by what are called `multi-black hole' solutions, which have horizons comprised of disconnected pieces, \cite{Elvang:2007rd, Elvang:2007hg, Elvang:2007hs} to name a few. The stationary sector of the large $D$ membrane paradigm contains membrane configurations that are dual to many of those. The membranes separated by distance $\gg \mathcal{O}(D^{-1})$ shouldn't interact with each-other, since in the corresponding black hole picture the black objects are away from each-others' gravitational influence. So one can easily construct multi-black objects, for example a black saturn \cite{Elvang:2007rd} can be obtained by simply putting a rotating black hole in the centre of a sufficiently thin black ring. However, this complete lack of interaction might lose some interesting features of compound objects like this, such as the rotational dragging of one object by another.

\section*{Acknowledgements}
I am very grateful to N. Kundu and A. Saha for the helpful discussions and vital inputs during writing this paper. I would also like to thank A. Bagchi, P. Banerjee, J. Bhattacharya, S. Bhattacharyya, Y. Dandekar, D. Das, K. Kolekar, S. Minwalla, S. Thakur and A. Yarom for the discussions and insights that were very useful for this project. I owe thanks to Technion and University of Haifa for the hospitality, where a significant initial part of this work was done. In the earlier stage this project was supported in part by an ISF excellence center grant 1989/14, a BSF grant 2016324 and by the Israel Science Foundation under grant 504/13. I would also like to acknowledge my debt to the people of India for their generous and steady support to research in the basic sciences.

\appendix
\section{Thermodynamic quantities in the units where $16\pi G =1$}\label{units}
In \cite{Mandlik:2018wnw}, the units were adopted so that $c=1$, $\hbar = 1$ and $G=1$. But in this note the author uses $c=1$, $\hbar = 1$ and $\color[rgb]{1.00,0.00,0.00}{16\pi G=1}$, because this removes some ugly overall factors of $\pi$ from the Einstein's field equations and the expression for stress tensor. So the first step is to put back the appropriate factors of $c$, $\hbar$ and $G$ in the expressions in \cite{Mandlik:2018wnw} by dimensional analysis, and then setting the new convention. But $D$ being large provides a shortcut. To appreciate this, let's first find out the dimensions of $G$. The gravitational potential due to a static, spherically symmetric, massive body in $D$ spacetime dimensions is given by
\begin{equation}
\Phi = -\frac{G M}{r^{D-3}}~.
\end{equation}
since $[\phi]=l^2t^{-2}$, $[G]$ = $m^{-1}l^{D-1}t^{-2}$. Thus even if we set $c=1$ and $\hbar =1$ early on, the hidden factors of $G$ can be easily exposed by looking at the behaviour of quantities with $D$. Setting $c$ and $\hbar$ to unity means setting $m^{-1}=t=l$, and so $[G] = l^{D-2}$. Notice that if $A$ is the area of the membrane, then $[A] = l^{D-2}$ as well.\\
The extrinsic quantities are obtained by integrating the densities over the area. Since neither the extrinsic quantities nor the densities have $l^D$ type behaviour, the dimension $l^{D-2}$ introduced by the integration has to be nullified by \underline{dividing} the densities by $G$. And then setting $16\pi G = 1$ results into simply multiplying the densities reported in \cite{Mandlik:2018wnw} by $16\pi$. The expressions for the intrinsic quantities, however, remain unchanged, and so do the stationary membrane equations.

So
\begin{equation}\label{Thermoq}
\begin{split}
T_{\mu\nu} &= \color[rgb]{1.00,0.00,0.00}{1}  K (1+Q^2)u_\mu u_\nu,\\
J_{\mu} &= \color[rgb]{1.00,0.00,0.00}{4\sqrt{2\pi}}  K Qu_{\mu},\\
J^{S}_{\mu} &= \color[rgb]{1.00,0.00,0.00}{4\pi} u_{\mu},
\end{split}
\end{equation}
while
\begin{equation}\label{intrinsic}
\begin{split}
T &= \frac{ K }{4\pi\gamma(1-Q^2)},\\
\mu &= \frac{Q}{2\sqrt{2\pi}\gamma}.
\end{split}
\end{equation}

\section{Free energy of static configurations in flat background at leading order in $D$}\label{grandcan}
For the static configurations in flat background, $\gamma =1$. So according to the stationary membrane equations, $Q$ and $ K $ are constant over the membrane. As computed in section \ref{thermosec},
\begin{equation}
\begin{split}
E &= \frac{D(1+\alpha^2)}{\beta(1-\alpha^2)}A\\
q &= \frac{4\sqrt{2\pi}D\alpha}{\beta(1-\alpha^2)}A,\\
S &= 4\pi A.
\end{split}
\end{equation}
where $A$ is the area of the membrane in this configuration. Also,
\begin{equation}\label{intrinsic2}
\begin{split}
T &= \frac{D}{4\pi\beta},\\
\mu &= \frac{\alpha}{2\sqrt{2\pi}}.
\end{split}
\end{equation}
So the free energy
\begin{equation}
 \begin{split}
 \mathcal{Q} &= E-TS-\mu q,\\
 &= \frac{DA}{\beta}\left(\frac{1+\alpha^2}{1-\alpha^2}-1-\frac{2\alpha^2}{1-\alpha^2}\right),\\
 &=0,
 \end{split}
\end{equation}
independent of the configuration. Hence, in a grand canonical ensemble, all the flat space static membrane configurations are equally likely.

\section{Christoffel Symbols}\label{christ}

\subsection{For the background spacetime metric in axisymmetric coordinates}

\begin{equation}\label{stchr}
\begin{split}
\Sigma^{\sigma}_{\phi\phi} &= -\sigma,\\
\Sigma^{\sigma}_{\theta\theta}&= - S\left(\sigma S+\frac{b}R\right),\\
\Sigma^{\sigma}_{ab} &=-\sigma C^2\hat{g}_{ab},\\
\Sigma^{\phi}_{\sigma\phi} &= \frac{1}\sigma,\\
\Sigma^{\theta}_{\sigma\theta} &= \frac{RS}{R\sigma S + b},\\
\Sigma^{a}_{\sigma b} &= \frac{\delta^a_b}{\sigma},\\
\Sigma^{\phi}_{\theta\theta}&= -\frac{C}{\sigma}\left(\sigma S+\frac{b}R\right),\\
\Sigma^{\phi}_{ab} &=SC\hat{g}_{ab},\\
\Sigma^{\theta}_{\phi\theta} &= \frac{R\sigma C}{R\sigma S+b},\\
\Sigma^{a}_{\phi b} &= \frac{-S}{C}\delta^a_b,\\
\Sigma^a_{bc} &= \hat{\Gamma}^a_{bc},
\end{split}
\end{equation}
where $\hat{\Gamma}^a_{bc}$ are the Christoffel symbols on the unit $S^{D-4}$.

\subsection{For the induced metric on the membrane in axisymmetric coordinates}

We split the nonzero Christoffel Symbols on the membrane in two categories. The ones that are $\mathcal{O}\left(\epsilon^0\right)$ are
\begin{equation}\label{memchr0}
\begin{split}
\Gamma^\phi_{\theta\theta} &= -\frac{1}R\left(C(RS+b)+\epsilon S(RS+b)\partial_\phi\delta\rho-\epsilon bC\delta\rho\right),\\
\Gamma^\phi_{ab} &= \left(SC-\epsilon C^2\partial_\phi\delta\rho\right)\hat{g}_{ab},\\
\Gamma^{\theta}_{\phi\theta} &= \frac{RC}{RS+b}+\epsilon\frac{RbC\delta\rho}{(RS+b)^2}+\epsilon\frac{RS\partial_\phi\delta\rho}{RS+b},\\
\Gamma^{a}_{\phi b} &= \left(\frac{-S}{C}+\epsilon\partial_\phi\delta\rho\right)\delta^a_b,\\
\Gamma^a_{bc} &= \hat{\Gamma}^a_{bc}+\epsilon\left(\delta^a_b\partial_c\delta\rho+\delta^a_c\partial_b\delta\rho-\hat{g}_{bc}\hat{g}^{ad}\partial_d\delta\rho\right).
\end{split}
\end{equation}

While the ones that are $\mathcal{O}(\epsilon)$ are
\begin{equation}\label{memchr1}
\begin{split}
\Gamma^t_{\phi\phi} &= \epsilon R^2 \partial_t\delta\rho,\\
\Gamma^t_{\theta\theta} &= \epsilon RS(RS+b) \partial_t\delta\rho,\\
\Gamma^t_{ab} &= \epsilon R^2C^2 \partial_t\delta\rho\hat{g}_{ab},\\
\Gamma^\phi_{t\phi} &= \epsilon\partial_t\delta\rho,\\
\Gamma^\theta_{t\theta} &= \epsilon\frac{RS}{RS+b}\partial_t\delta\rho,\\
\Gamma^a_{tb} &= \epsilon\partial_t\delta\rho\delta^a_b,\\
\Gamma^\phi_{\phi\phi} &= \epsilon\partial_\phi\delta\rho,\\
\Gamma^\phi_{\theta\phi} &= \epsilon\partial_\theta\delta\rho,\\
\Gamma^\phi_{a\phi} &= \epsilon\partial_a\delta\rho,\\
\Gamma^\theta_{\phi\phi} &= \frac{-\epsilon R^2\partial_\theta\delta\rho}{(RS+b)^2},\\
\Gamma^a_{\phi\phi} &= -\epsilon\hat{g}^{ab}\frac{\partial_b\delta\rho}{C^2},\\
\Gamma^{\theta}_{\theta\theta} &= \epsilon\frac{RS}{RS+b}\partial_\theta\delta\rho,\\
\Gamma^{a}_{\theta\theta} &=-\epsilon S\frac{RS+b}{RC^2}\hat{g}^{ab}\partial_b\delta\rho,\\
\Gamma^{\theta}_{a\theta} &=\epsilon \frac{RS}{RS+b}\partial_a\delta\rho,\\
\Gamma^\theta_{ab} &= -\epsilon\frac{R^2C^2}{(RS+b)^2}\hat{g}_{ab}\partial_\theta\delta\rho,\\
\Gamma^a_{\theta b} &= \epsilon~\partial_\theta\delta\rho~\delta^a_b.\\
\end{split}
\end{equation}
The unperturbed Christoffel symbols in \eqref{memchr1} vanish.

\section{Black ring as $S^1 \times S^{D-3}$}\label{SD-3}

In the section \ref{LQNM} the membrane is defined by first assigning the axisymmetric coordinate system $(t,\sigma, \phi,\theta,\{\chi^a\})$ to the background flat spacetime and then setting $\sigma =1$ to get the embedding of the membrane in it. In constructing the axisymmetric coordinates, one plane is separated from the constant time slice of the spacetime (referred to as space), giving the space a $R^2 \times R^{D-3}$ structure. Thus  the angles $\{\chi^a\}$, which generate the $SO(D-3)$ symmetry of the $R^{D-3}$ part generate $S^{D-4}$. Such a construction doesn't immediately reveal the $S^{D-3}$ part of the black ring's topology. In this appendix this $S^{D-4}$ is shown to be a section of $S^{D-3}$.\\

Consider a $n+1$ dimensional unit sphere. One can define an orthogonal coordinate system $(\Theta^{\Xi})$ on it, where $\Xi$ takes integer values from $1$ to $n+1$. The range of these coordinates is
$$-\frac{\pi}2\leq\Theta^{\Xi}\leq \frac{\pi}2~~~\text{for}~2\leq \Xi \leq n+1,~~~ -\pi<\Theta^1\leq \pi,$$
where $\Theta^1 = -\pi$ is identified with $\Theta^1 = \pi$. In this coordinate system the metric $d\Omega^2_{n+1}$ on $S^{n+1}$ is obtained recursively as
\begin{equation}
\begin{split}
d\Omega^2_{(\Xi+1)} &= (d\Theta^{\Xi+1})^2 + (\cos{\Theta^{\Xi+1}})^2d\Omega^2_{(\Xi)}~~~\text{for}~1\leq \Xi \leq n,\\
d\Omega^2_{(1)} &= (d\Theta^{1})^2.
\end{split}
\end{equation}
Now in this framework, if we could identify $n=D-4$, $\Theta^{D-3} = \phi$ and $\Theta^a = \chi^a$ for all $1\leq a \leq {D-4}$, we would have shown that coordinates $(\phi,\{\chi^a\})$ describe an $S^{D-3}$ topology. For this, the ranges of the identified coordinates have to match, and the singularities, if any, also need to be identical. The range of $\phi$ is $[-\frac{\pi}2,\frac{\pi}2]$ which is the same as that of $\Theta^{D-3}$, and since $\{\chi^a\}$ are the angles on $S^{D-4}$, by a similar construction as above, we can choose them so that 
$$-\frac{\pi}2\leq\chi^{a}\leq \frac{\pi}2~~~\text{for}~2\leq a \leq D-4,~~~ -\pi<\chi^1\leq \pi,$$
so they have the desired range.

Now let's match the singularities. On $S^{D-3}$ constructed above the size of the $S^{D-4}$ spanned by $\Theta^\Xi,~~1\leq\Xi\leq (D-4)$ vanishes at the two `poles', $\Theta^{D-3} = \pm \frac{\pi}2$. Looking at the constant time and constant $\theta$ section of the unperturbed black ring metric \eqref{indmet} ($\epsilon = 0$), we get the restriction of the metric to be $R^2(d\phi^2 + \cos(\phi)^2 d\Omega_{D-4}^2)$, so at the poles $\phi = \pm \frac{\pi}2$ the size of the $S^{D-4}$ vanishes. Thus this section has an $S^{D-3}$ topology. In fact, the metric says that it {\it is} exactly a sphere, of radius $R$.

Now let's write the divergence and laplacians on a unit $S^{D-3}$ ($R=1$) in terms of $(\phi,\{\chi^a\})$ as required by \ref{lightQNM}. First, the Christoffel symbols on $S^{D-3}$ can be obtained from \eqref{memchr0} (after setting $\epsilon = 0$) because of the block diagonal nature of the metric \eqref{indmet}:

\begin{equation}\label{SD-3Chr}
\begin{split}
\Gamma^\phi_{ab} &= SC\hat{g}_{ab},\\
\Gamma^{a}_{\phi b} &= \frac{-S}{C}\delta^a_b,\\
\Gamma^a_{bc} &= \hat{\Gamma}^a_{bc}.
\end{split}
\end{equation}

The divergence of a vector $V^\Xi \equiv (V^\phi, \{V^a\})$ is, to the order $D$,
\begin{equation}\label{Divv}
\nabla\cdot V \equiv \nabla_\Xi V^\Xi = \hat\nabla_aV^a-\frac{DS}{C}V^\phi + \mathcal{O}(D^0).
\end{equation}
The velocity fluctuations, in the absence of shape fluctuations, are divergenceless on the $S^{D-3}$ too, according to \eqref{divucond}. So they should be able to be decomposed into the vector spherical harmonics on $S^{D-3}$.

The laplacian on $S^{D-3}$ acting on a scalar function becomes to the leading order
\begin{equation}\label{LapS}
\nabla^2 \mathcal{S} = \left(\frac{\hat{\nabla}^2S}{C^2}-\frac{DS}{C}\partial_\phi\right)\mathcal{S} \equiv D\mathcal{D}~\mathcal{S},
\end{equation}
while the Laplacian acting on vector $V^\Xi$ to the leading order is
\begin{equation}\label{LapV}
\begin{split}
\nabla^2 V^\phi &=  \left(\frac{\hat{\nabla}^2S}{C^2}-\frac{DS}{C}\partial_\phi-\frac{DS^2}{C^2}\right)V^{\phi} + 2\frac{S}{C}\hat\nabla_aV^a,\\
\nabla^2 V^a &=  \left(\frac{\hat{\nabla}^2S}{C^2}-\frac{DS}{C}\partial_\phi+\frac{DS^2}{C^2}\right)V^a.
\end{split}
\end{equation}
If $V$ is a vector spherical harmonic, $\nabla\cdot V = 0$. Then using \eqref{Divv},
\begin{equation}\label{LapdivV}
\begin{split}
\nabla^2 V^\phi &=  \left(\frac{\hat{\nabla}^2S}{C^2}-\frac{DS}{C}\partial_\phi+\frac{DS^2}{C^2}\right)V^{\phi}\\
\nabla^2 V^a &=  \left(\frac{\hat{\nabla}^2S}{C^2}-\frac{DS}{C}\partial_\phi-\frac{DS^2}{C^2}\right)V^a
\end{split}
\end{equation}
Compactly written,
\begin{equation}
\begin{split}
\nabla^2 V^\Xi &=  \left(\frac{\hat{\nabla}^2S}{C^2}-\frac{DS}{C}\partial_\phi+\frac{DS^2}{C^2}\right)V^{\Xi}\\
&\equiv D\mathcal{D}_V~V^{\Xi}
\end{split}
\end{equation}

\section{Solving subleading order membrane equations for axisymmetric stationary black ring configuration}\label{subrot}

In the $(t,r,\theta,s,\{\chi^a\})$ coordinate system, the stationary axisymmetric membrane is given by
\begin{equation}
s^2 = 2g(r),
\end{equation}
and solving the membrane equation for $g(r)$ gives the desired membrane configuration. As discussed in the end of the section \ref{rotBRsubl}, the stationary, axisymmetric uncharged membrane equation takes the form up to $\mathcal{O}\left(D^0\right)$
\begin{equation}\label{Ksubeq}
K = \frac{D-4}{\beta}\left(1-\frac{\tilde{\omega}^2r^2}{D-4}\right).
\end{equation}
\footnote{Note that \eqref{Ksubeq} looks different from the similar equation at the end of \ref{rotBRsubl} because $D$ has been replaced by $D-4$. The analysis in that section was schematic, while the analysis here is more careful.}
First we write K in terms of the shape function $\eqref{shapesubl}$. The unit normal to the membrane $s^2 = 2g(r)$ has components
\begin{equation}\label{norm}
n_s = \frac{s}{\sqrt{s^2+(g')^2}},~~~n_r = \frac{g'}{\sqrt{s^2+(g')^2}},
\end{equation}
where $g' \equiv \frac{dg}{dr}$. Keep in mind that on the membrane $s^2 = 2g$.\\
Now the trace of extrinsic curvature, $K$ is given by
\begin{equation}
K = \nabla_A n^A = \frac{1}{\sqrt{-G}}\partial_A \left(\sqrt{-G}G^{AB}n_B\right).
\end{equation}
In this coordinate system, $\sqrt{-G} = rs^{D-4}\sqrt{\hat{g}}$. $\hat{g}$ is the volume element of unit $S^{D-4}$. Since the metric is diagonal and $n_A$ doesn't have any $\chi^a$ dependence or components along them, $\hat{g}$ factor drops out and we get
\begin{equation}\label{Ksubgen}
K = (D-4)\frac{n_s}{s} + \frac{n_r}{r}+\partial_rn_r+\partial_sn_s.
\end{equation}
From \eqref{norm}, and denoting $\frac{1}{D-4}$ by $\epsilon$, we get
\begin{equation}\label{K1}
\begin{split}
\partial_rn_r+\partial_sn_s &= \frac{(g')^2-2gg''}{\left(2g+(g')^2\right)^{\frac{3}2}}~,\\
&= \frac{1}{\beta}\left[1+\frac{\epsilon}{\beta^2}\left(\left[(r-b)^2-\beta^2\right]\mathfrak{g}''+(r-b)\mathfrak{g}'-\mathfrak{g}\right)\right],
\end{split}
\end{equation}
and
\begin{equation}\label{K2}
(D-4)\frac{n_s}{s} = \frac{1}{\epsilon\beta}\left[1-\frac{\epsilon}{\beta^2}\left(\mathfrak{g} - (r-b)\mathfrak{g}'\right)\right],
\end{equation}
and
\begin{equation}\label{K3}
\frac{n_r}{r} = \frac{1}{\beta}\left(1-\frac{b}{r}\right)\left[1-\frac{\epsilon}{\beta^2}\left(\mathfrak{g}-(r-b)\mathfrak{g}'\right)\right]-\frac{\epsilon}{\beta r}\mathfrak{g}'~.
\end{equation}
Evaluating $K$ from \eqref{Ksubgen} by substituting \eqref{K1}, \eqref{K2} and \eqref{K3}, the equation \eqref{Ksubeq} at $\mathcal{O}(\epsilon^{-1})$ becomes the leading order equation which is solved by the leading order solution, while at the order $\epsilon^0$ it reads:
\begin{equation}
\frac{1}{\beta}\left(2-\frac{b}r\right)-\frac{1}{\beta^2}\left[\mathfrak{g}-(r-b)\mathfrak{g}'\right].
\end{equation}
This is a first order ordinary differential equation which can be solved exactly
\begin{equation}
\mathfrak{g} = \beta^2\left[\tilde{\omega}^2\left(r(r-b)+b^2\right)+\frac{r-b}b\ln{r}-(1-2\tilde{\omega}^2b^2)\frac{r-b}b\ln{(r-b)}+c(r-b)\right],
\end{equation}
where $c$ is the integration constant. For this solution to be regular over the ring, the coefficient of the irregularity $\ln{(r-b)}$ has to vanish. this gives the condition
\begin{equation}\label{rotcond}
\tilde{\omega} = \frac{1}{\sqrt{2}~b}~.
\end{equation}

\newpage
\bibliographystyle{JHEP}
\bibliography{Stationary}

\providecommand{\href}[2]{#2}\begingroup\raggedright\begin{thebibliography}{10}

\bibitem{Hawking:1973uf}
S.~Hawking and G.~Ellis, \emph{{The Large Scale Structure of Space-Time}},
  Cambridge Monographs on Mathematical Physics. Cambridge University Press, 2,
  2011,
  \href{https://doi.org/10.1017/CBO9780511524646}{10.1017/CBO9780511524646}.

\bibitem{Cai:2001su}
M.-l. Cai and G.~J. Galloway, \emph{{On the Topology and area of higher
  dimensional black holes}},
  \href{https://doi.org/10.1088/0264-9381/18/14/308}{\emph{Class. Quant. Grav.}
  {\bfseries 18} (2001) 2707}
  [\href{https://arxiv.org/abs/hep-th/0102149}{{\ttfamily hep-th/0102149}}].

\bibitem{Myers:1986un}
R.~C. Myers and M.~Perry, \emph{{Black Holes in Higher Dimensional
  Space-Times}},
  \href{https://doi.org/10.1016/0003-4916(86)90186-7}{\emph{Annals Phys.}
  {\bfseries 172} (1986) 304}.

\bibitem{Myers:2011yc}
R.~C. Myers, \emph{{Myers--Perry black holes}}, pp.~101--133.
\newblock 2012.
\newblock \href{https://arxiv.org/abs/1111.1903}{{\ttfamily 1111.1903}}.

\bibitem{Emparan:2001wk}
R.~Emparan and H.~S. Reall, \emph{{Generalized Weyl solutions}},
  \href{https://doi.org/10.1103/PhysRevD.65.084025}{\emph{Phys. Rev. D}
  {\bfseries 65} (2002) 084025}
  [\href{https://arxiv.org/abs/hep-th/0110258}{{\ttfamily hep-th/0110258}}].

\bibitem{Emparan:2001wn}
R.~Emparan and H.~S. Reall, \emph{{A Rotating black ring solution in
  five-dimensions}},
  \href{https://doi.org/10.1103/PhysRevLett.88.101101}{\emph{Phys. Rev. Lett.}
  {\bfseries 88} (2002) 101101}
  [\href{https://arxiv.org/abs/hep-th/0110260}{{\ttfamily hep-th/0110260}}].

\bibitem{Emparan:2006mm}
R.~Emparan and H.~S. Reall, \emph{{Black Rings}},
  \href{https://doi.org/10.1088/0264-9381/23/20/R01}{\emph{Class. Quant. Grav.}
  {\bfseries 23} (2006) R169}
  [\href{https://arxiv.org/abs/hep-th/0608012}{{\ttfamily hep-th/0608012}}].

\bibitem{Chervonyi:2015uua}
Y.~Chervonyi, \emph{{Towards higher dimensional black rings}},
  \href{https://doi.org/10.1103/PhysRevD.92.124037}{\emph{Phys. Rev. D}
  {\bfseries 92} (2015) 124037}
  [\href{https://arxiv.org/abs/1510.06041}{{\ttfamily 1510.06041}}].

\bibitem{Emparan:2007wm}
R.~Emparan, T.~Harmark, V.~Niarchos, N.~A. Obers and M.~J. Rodriguez,
  \emph{{The Phase Structure of Higher-Dimensional Black Rings and Black
  Holes}}, \href{https://doi.org/10.1088/1126-6708/2007/10/110}{\emph{JHEP}
  {\bfseries 10} (2007) 110} [\href{https://arxiv.org/abs/0708.2181}{{\ttfamily
  0708.2181}}].

\bibitem{Caldarelli:2008pz}
M.~M. Caldarelli, R.~Emparan and M.~J. Rodriguez, \emph{{Black Rings in
  (Anti)-deSitter space}},
  \href{https://doi.org/10.1088/1126-6708/2008/11/011}{\emph{JHEP} {\bfseries
  11} (2008) 011} [\href{https://arxiv.org/abs/0806.1954}{{\ttfamily
  0806.1954}}].

\bibitem{Kleihaus:2012xh}
B.~Kleihaus, J.~Kunz and E.~Radu, \emph{{Black rings in six dimensions}},
  \href{https://doi.org/10.1016/j.physletb.2012.11.015}{\emph{Phys. Lett. B}
  {\bfseries 718} (2013) 1073}
  [\href{https://arxiv.org/abs/1205.5437}{{\ttfamily 1205.5437}}].

\bibitem{Emparan:2020inr}
R.~Emparan and C.~P. Herzog, \emph{{The Large D Limit of Einstein's
  Equations}},  \href{https://arxiv.org/abs/2003.11394}{{\ttfamily
  2003.11394}}.

\bibitem{Emparan:2013moa}
R.~Emparan, R.~Suzuki and K.~Tanabe, \emph{{The large D limit of General
  Relativity}}, \href{https://doi.org/10.1007/JHEP06(2013)009}{\emph{JHEP}
  {\bfseries 06} (2013) 009} [\href{https://arxiv.org/abs/1302.6382}{{\ttfamily
  1302.6382}}].

\bibitem{Emparan:2013xia}
R.~Emparan, D.~Grumiller and K.~Tanabe, \emph{{Large-D gravity and low-D
  strings}}, \href{https://doi.org/10.1103/PhysRevLett.110.251102}{\emph{Phys.
  Rev. Lett.} {\bfseries 110} (2013) 251102}
  [\href{https://arxiv.org/abs/1303.1995}{{\ttfamily 1303.1995}}].

\bibitem{Emparan:2013oza}
R.~Emparan and K.~Tanabe, \emph{{Holographic superconductivity in the large D
  expansion}}, \href{https://doi.org/10.1007/JHEP01(2014)145}{\emph{JHEP}
  {\bfseries 01} (2014) 145} [\href{https://arxiv.org/abs/1312.1108}{{\ttfamily
  1312.1108}}].

\bibitem{Emparan:2014cia}
R.~Emparan and K.~Tanabe, \emph{{Universal quasinormal modes of large D black
  holes}}, \href{https://doi.org/10.1103/PhysRevD.89.064028}{\emph{Phys. Rev.
  D} {\bfseries 89} (2014) 064028}
  [\href{https://arxiv.org/abs/1401.1957}{{\ttfamily 1401.1957}}].

\bibitem{Emparan:2014jca}
R.~Emparan, R.~Suzuki and K.~Tanabe, \emph{{Instability of rotating black
  holes: large D analysis}},
  \href{https://doi.org/10.1007/JHEP06(2014)106}{\emph{JHEP} {\bfseries 06}
  (2014) 106} [\href{https://arxiv.org/abs/1402.6215}{{\ttfamily 1402.6215}}].

\bibitem{Emparan:2014aba}
R.~Emparan, R.~Suzuki and K.~Tanabe, \emph{{Decoupling and non-decoupling
  dynamics of large D black holes}},
  \href{https://doi.org/10.1007/JHEP07(2014)113}{\emph{JHEP} {\bfseries 07}
  (2014) 113} [\href{https://arxiv.org/abs/1406.1258}{{\ttfamily 1406.1258}}].

\bibitem{Emparan:2015rva}
R.~Emparan, R.~Suzuki and K.~Tanabe, \emph{{Quasinormal modes of (Anti-)de
  Sitter black holes in the 1/D expansion}},
  \href{https://doi.org/10.1007/JHEP04(2015)085}{\emph{JHEP} {\bfseries 04}
  (2015) 085} [\href{https://arxiv.org/abs/1502.02820}{{\ttfamily
  1502.02820}}].

\bibitem{Emparan:2015hwa}
R.~Emparan, T.~Shiromizu, R.~Suzuki, K.~Tanabe and T.~Tanaka, \emph{{Effective
  theory of Black Holes in the 1/D expansion}},
  \href{https://doi.org/10.1007/JHEP06(2015)159}{\emph{JHEP} {\bfseries 06}
  (2015) 159} [\href{https://arxiv.org/abs/1504.06489}{{\ttfamily
  1504.06489}}].

\bibitem{Suzuki:2015iha}
R.~Suzuki and K.~Tanabe, \emph{{Stationary black holes: Large $D$ analysis}},
  \href{https://doi.org/10.1007/JHEP09(2015)193}{\emph{JHEP} {\bfseries 09}
  (2015) 193} [\href{https://arxiv.org/abs/1505.01282}{{\ttfamily
  1505.01282}}].

\bibitem{Suzuki:2015axa}
R.~Suzuki and K.~Tanabe, \emph{{Non-uniform black strings and the critical
  dimension in the $1/D$ expansion}},
  \href{https://doi.org/10.1007/JHEP10(2015)107}{\emph{JHEP} {\bfseries 10}
  (2015) 107} [\href{https://arxiv.org/abs/1506.01890}{{\ttfamily
  1506.01890}}].

\bibitem{Emparan:2015gva}
R.~Emparan, R.~Suzuki and K.~Tanabe, \emph{{Evolution and End Point of the
  Black String Instability: Large D Solution}},
  \href{https://doi.org/10.1103/PhysRevLett.115.091102}{\emph{Phys. Rev. Lett.}
  {\bfseries 115} (2015) 091102}
  [\href{https://arxiv.org/abs/1506.06772}{{\ttfamily 1506.06772}}].

\bibitem{Tanabe:2015hda}
K.~Tanabe, \emph{{Black rings at large D}},
  \href{https://doi.org/10.1007/JHEP02(2016)151}{\emph{JHEP} {\bfseries 02}
  (2016) 151} [\href{https://arxiv.org/abs/1510.02200}{{\ttfamily
  1510.02200}}].

\bibitem{Tanabe:2015isb}
K.~Tanabe, \emph{{Instability of the de Sitter Reissner--Nordstrom black hole
  in the $1/D$ expansion}},
  \href{https://doi.org/10.1088/0264-9381/33/12/125016}{\emph{Class. Quant.
  Grav.} {\bfseries 33} (2016) 125016}
  [\href{https://arxiv.org/abs/1511.06059}{{\ttfamily 1511.06059}}].

\bibitem{Andrade:2015hpa}
T.~Andrade, S.~A. Gentle and B.~Withers, \emph{{Drude in D major}},
  \href{https://doi.org/10.1007/JHEP06(2016)134}{\emph{JHEP} {\bfseries 06}
  (2016) 134} [\href{https://arxiv.org/abs/1512.06263}{{\ttfamily
  1512.06263}}].

\bibitem{Emparan:2016sjk}
R.~Emparan, K.~Izumi, R.~Luna, R.~Suzuki and K.~Tanabe, \emph{{Hydro-elastic
  Complementarity in Black Branes at large D}},
  \href{https://doi.org/10.1007/JHEP06(2016)117}{\emph{JHEP} {\bfseries 06}
  (2016) 117} [\href{https://arxiv.org/abs/1602.05752}{{\ttfamily
  1602.05752}}].

\bibitem{Tanabe:2016pjr}
K.~Tanabe, \emph{{Elastic instability of black rings at large D}},
  \href{https://arxiv.org/abs/1605.08116}{{\ttfamily 1605.08116}}.

\bibitem{Tanabe:2016opw}
K.~Tanabe, \emph{{Charged rotating black holes at large D}},
  \href{https://arxiv.org/abs/1605.08854}{{\ttfamily 1605.08854}}.

\bibitem{Andrade:2018zeb}
T.~Andrade, C.~Pantelidou and B.~Withers, \emph{{Large D holography with metric
  deformations}}, \href{https://doi.org/10.1007/JHEP09(2018)138}{\emph{JHEP}
  {\bfseries 09} (2018) 138}
  [\href{https://arxiv.org/abs/1806.00306}{{\ttfamily 1806.00306}}].

\bibitem{Andrade:2018nsz}
T.~Andrade, R.~Emparan and D.~Licht, \emph{{Rotating black holes and black bars
  at large D}}, \href{https://doi.org/10.1007/JHEP09(2018)107}{\emph{JHEP}
  {\bfseries 09} (2018) 107}
  [\href{https://arxiv.org/abs/1807.01131}{{\ttfamily 1807.01131}}].

\bibitem{Emparan:2019obu}
R.~Emparan and R.~Suzuki, \emph{{Topology-changing horizons at large D as Ricci
  flows}}, \href{https://doi.org/10.1007/JHEP07(2019)094}{\emph{JHEP}
  {\bfseries 07} (2019) 094}
  [\href{https://arxiv.org/abs/1905.01062}{{\ttfamily 1905.01062}}].

\bibitem{Andrade:2019edf}
T.~Andrade, R.~Emparan, D.~Licht and R.~Luna, \emph{{Black hole collisions,
  instabilities, and cosmic censorship violation at large $D$}},
  \href{https://doi.org/10.1007/JHEP09(2019)099}{\emph{JHEP} {\bfseries 09}
  (2019) 099} [\href{https://arxiv.org/abs/1908.03424}{{\ttfamily
  1908.03424}}].

\bibitem{Licht:2020odx}
D.~Licht, R.~Luna and R.~Suzuki, \emph{{Black Ripples, Flowers and Dumbbells at
  large $D$}}, \href{https://doi.org/10.1007/JHEP04(2020)108}{\emph{JHEP}
  {\bfseries 04} (2020) 108}
  [\href{https://arxiv.org/abs/2002.07813}{{\ttfamily 2002.07813}}].

\bibitem{Chen:2015fuf}
B.~Chen, Z.-Y. Fan, P.~Li and W.~Ye, \emph{{Quasinormal modes of Gauss-Bonnet
  black holes at large D}},
  \href{https://doi.org/10.1007/JHEP01(2016)085}{\emph{JHEP} {\bfseries 01}
  (2016) 085} [\href{https://arxiv.org/abs/1511.08706}{{\ttfamily
  1511.08706}}].

\bibitem{Chen:2016fuy}
B.~Chen and P.-C. Li, \emph{{Instability of Charged Gauss-Bonnet Black Hole in
  de Sitter Spacetime at Large $D$}},
  \href{https://arxiv.org/abs/1607.04713}{{\ttfamily 1607.04713}}.

\bibitem{Chen:2017wpf}
B.~Chen, P.-C. Li and Z.-z. Wang, \emph{{Charged Black Rings at large D}},
  \href{https://doi.org/10.1007/JHEP04(2017)167}{\emph{JHEP} {\bfseries 04}
  (2017) 167} [\href{https://arxiv.org/abs/1702.00886}{{\ttfamily
  1702.00886}}].

\bibitem{Chen:2017hwm}
B.~Chen and P.-C. Li, \emph{{Static Gauss-Bonnet Black Holes at Large $D$}},
  \href{https://doi.org/10.1007/JHEP05(2017)025}{\emph{JHEP} {\bfseries 05}
  (2017) 025} [\href{https://arxiv.org/abs/1703.06381}{{\ttfamily
  1703.06381}}].

\bibitem{Chen:2017rxa}
B.~Chen, P.-C. Li and C.-Y. Zhang, \emph{{Einstein-Gauss-Bonnet Black Strings
  at Large $D$}}, \href{https://doi.org/10.1007/JHEP10(2017)123}{\emph{JHEP}
  {\bfseries 10} (2017) 123}
  [\href{https://arxiv.org/abs/1707.09766}{{\ttfamily 1707.09766}}].

\bibitem{Chen:2018nbh}
B.~Chen, P.-C. Li, Y.~Tian and C.-Y. Zhang, \emph{{Holographic Turbulence in
  Einstein-Gauss-Bonnet Gravity at Large $D$}},
  \href{https://doi.org/10.1007/JHEP01(2019)156}{\emph{JHEP} {\bfseries 01}
  (2019) 156} [\href{https://arxiv.org/abs/1804.05182}{{\ttfamily
  1804.05182}}].

\bibitem{Chen:2018vbv}
B.~Chen, P.-C. Li and C.-Y. Zhang, \emph{{Einstein-Gauss-Bonnet Black Rings at
  Large $D$}}, \href{https://doi.org/10.1007/JHEP07(2018)067}{\emph{JHEP}
  {\bfseries 07} (2018) 067}
  [\href{https://arxiv.org/abs/1805.03345}{{\ttfamily 1805.03345}}].

\bibitem{Li:2019bqc}
P.-C. Li, C.-Y. Zhang and B.~Chen, \emph{{The Fate of Instability of de Sitter
  Black Holes at Large $D$}},
  \href{https://doi.org/10.1007/JHEP11(2019)042}{\emph{JHEP} {\bfseries 11}
  (2019) 042} [\href{https://arxiv.org/abs/1909.02685}{{\ttfamily
  1909.02685}}].

\bibitem{Guo:2019pte}
M.~Guo, P.-C. Li and B.~Chen, \emph{{Photon Emission Near Myers-Perry Black
  Holes in the Large Dimension Limit}},
  \href{https://doi.org/10.1103/PhysRevD.101.024054}{\emph{Phys. Rev. D}
  {\bfseries 101} (2020) 024054}
  [\href{https://arxiv.org/abs/1911.08814}{{\ttfamily 1911.08814}}].

\bibitem{Herzog:2016hob}
C.~P. Herzog, M.~Spillane and A.~Yarom, \emph{{The holographic dual of a
  Riemann problem in a large number of dimensions}},
  \href{https://doi.org/10.1007/JHEP08(2016)120}{\emph{JHEP} {\bfseries 08}
  (2016) 120} [\href{https://arxiv.org/abs/1605.01404}{{\ttfamily
  1605.01404}}].

\bibitem{Rozali:2016yhw}
M.~Rozali and A.~Vincart-Emard, \emph{{On Brane Instabilities in the Large $D$
  Limit}}, \href{https://doi.org/10.1007/JHEP08(2016)166}{\emph{JHEP}
  {\bfseries 08} (2016) 166}
  [\href{https://arxiv.org/abs/1607.01747}{{\ttfamily 1607.01747}}].

\bibitem{Rozali:2017bll}
M.~Rozali, E.~Sabag and A.~Yarom, \emph{{Holographic Turbulence in a Large
  Number of Dimensions}},
  \href{https://doi.org/10.1007/JHEP04(2018)065}{\emph{JHEP} {\bfseries 04}
  (2018) 065} [\href{https://arxiv.org/abs/1707.08973}{{\ttfamily
  1707.08973}}].

\bibitem{Herzog:2017qwp}
C.~P. Herzog and Y.~Kim, \emph{{The Large Dimension Limit of a Small Black Hole
  Instability in Anti-de Sitter Space}},
  \href{https://doi.org/10.1007/JHEP02(2018)167}{\emph{JHEP} {\bfseries 02}
  (2018) 167} [\href{https://arxiv.org/abs/1711.04865}{{\ttfamily
  1711.04865}}].

\bibitem{Rozali:2018yrv}
M.~Rozali and B.~Way, \emph{{Gravitating scalar stars in the large D limit}},
  \href{https://doi.org/10.1007/JHEP11(2018)106}{\emph{JHEP} {\bfseries 11}
  (2018) 106} [\href{https://arxiv.org/abs/1807.10283}{{\ttfamily
  1807.10283}}].

\bibitem{Casalderrey-Solana:2018uag}
J.~Casalderrey-Solana, C.~P. Herzog and B.~Meiring, \emph{{Holographic Bjorken
  Flow at Large-$D$}},
  \href{https://doi.org/10.1007/JHEP01(2019)181}{\emph{JHEP} {\bfseries 01}
  (2019) 181} [\href{https://arxiv.org/abs/1810.02314}{{\ttfamily
  1810.02314}}].

\bibitem{Sadhu:2016ynd}
A.~Sadhu and V.~Suneeta, \emph{{Nonspherically symmetric black string
  perturbations in the large dimension limit}},
  \href{https://doi.org/10.1103/PhysRevD.93.124002}{\emph{Phys. Rev. D}
  {\bfseries 93} (2016) 124002}
  [\href{https://arxiv.org/abs/1604.00595}{{\ttfamily 1604.00595}}].

\bibitem{Sadhu:2018zyh}
A.~Sadhu and V.~Suneeta, \emph{{Schwarzschild-Tangherlini quasinormal modes at
  large $D$ revisited}},  \href{https://arxiv.org/abs/1806.04888}{{\ttfamily
  1806.04888}}.

\bibitem{Sadhu:2018asi}
A.~Sadhu and V.~Suneeta, \emph{{Study of Semiclassical Instability of the
  Schwarzschild AdS Black Hole in the Large $D$ Limit}},
  \href{https://doi.org/10.1088/1361-6382/ab1d69}{\emph{Class. Quant. Grav.}
  {\bfseries 36} (2019) 115005}
  [\href{https://arxiv.org/abs/1812.11742}{{\ttfamily 1812.11742}}].

\bibitem{Bhattacharyya:2015dva}
S.~Bhattacharyya, A.~De, S.~Minwalla, R.~Mohan and A.~Saha, \emph{{A membrane
  paradigm at large D}},
  \href{https://doi.org/10.1007/JHEP04(2016)076}{\emph{JHEP} {\bfseries 04}
  (2016) 076} [\href{https://arxiv.org/abs/1504.06613}{{\ttfamily
  1504.06613}}].

\bibitem{Bhattacharyya:2015fdk}
S.~Bhattacharyya, M.~Mandlik, S.~Minwalla and S.~Thakur, \emph{{A Charged
  Membrane Paradigm at Large D}},
  \href{https://doi.org/10.1007/JHEP04(2016)128}{\emph{JHEP} {\bfseries 04}
  (2016) 128} [\href{https://arxiv.org/abs/1511.03432}{{\ttfamily
  1511.03432}}].

\bibitem{Dandekar:2016fvw}
Y.~Dandekar, A.~De, S.~Mazumdar, S.~Minwalla and A.~Saha, \emph{{The large D
  black hole Membrane Paradigm at first subleading order}},
  \href{https://doi.org/10.1007/JHEP12(2016)113}{\emph{JHEP} {\bfseries 12}
  (2016) 113} [\href{https://arxiv.org/abs/1607.06475}{{\ttfamily
  1607.06475}}].

\bibitem{Dandekar:2016jrp}
Y.~Dandekar, S.~Mazumdar, S.~Minwalla and A.~Saha, \emph{{Unstable `black
  branes' from scaled membranes at large $D$}},
  \href{https://doi.org/10.1007/JHEP12(2016)140}{\emph{JHEP} {\bfseries 12}
  (2016) 140} [\href{https://arxiv.org/abs/1609.02912}{{\ttfamily
  1609.02912}}].

\bibitem{Bhattacharyya:2016nhn}
S.~Bhattacharyya, A.~K. Mandal, M.~Mandlik, U.~Mehta, S.~Minwalla, U.~Sharma
  et~al., \emph{{Currents and Radiation from the large $D$ Black Hole
  Membrane}}, \href{https://doi.org/10.1007/JHEP05(2017)098}{\emph{JHEP}
  {\bfseries 05} (2017) 098}
  [\href{https://arxiv.org/abs/1611.09310}{{\ttfamily 1611.09310}}].

\bibitem{Bhattacharyya:2017hpj}
S.~Bhattacharyya, P.~Biswas, B.~Chakrabarty, Y.~Dandekar and A.~Dinda,
  \emph{{The large D black hole dynamics in AdS/dS backgrounds}},
  \href{https://doi.org/10.1007/JHEP10(2018)033}{\emph{JHEP} {\bfseries 10}
  (2018) 033} [\href{https://arxiv.org/abs/1704.06076}{{\ttfamily
  1704.06076}}].

\bibitem{Dandekar:2017aiv}
Y.~Dandekar, S.~Kundu, S.~Mazumdar, S.~Minwalla, A.~Mishra and A.~Saha,
  \emph{{An Action for and Hydrodynamics from the improved Large D membrane}},
  \href{https://doi.org/10.1007/JHEP09(2018)137}{\emph{JHEP} {\bfseries 09}
  (2018) 137} [\href{https://arxiv.org/abs/1712.09400}{{\ttfamily
  1712.09400}}].

\bibitem{Bhattacharyya:2018szu}
S.~Bhattacharyya, P.~Biswas and Y.~Dandekar, \emph{{Black holes in presence of
  cosmological constant: second order in $ \frac{1}{D} $}},
  \href{https://doi.org/10.1007/JHEP10(2018)171}{\emph{JHEP} {\bfseries 10}
  (2018) 171} [\href{https://arxiv.org/abs/1805.00284}{{\ttfamily
  1805.00284}}].

\bibitem{Mandlik:2018wnw}
M.~Mandlik and S.~Thakur, \emph{{Stationary Solutions from the Large D Membrane
  Paradigm}}, \href{https://doi.org/10.1007/JHEP11(2018)026}{\emph{JHEP}
  {\bfseries 11} (2018) 026}
  [\href{https://arxiv.org/abs/1806.04637}{{\ttfamily 1806.04637}}].

\bibitem{Saha:2018elg}
A.~Saha, \emph{{The large D Membrane Paradigm For Einstein-Gauss-Bonnet
  Gravity}}, \href{https://doi.org/10.1007/JHEP01(2019)028}{\emph{JHEP}
  {\bfseries 01} (2019) 028}
  [\href{https://arxiv.org/abs/1806.05201}{{\ttfamily 1806.05201}}].

\bibitem{Kundu:2018dvx}
S.~Kundu and P.~Nandi, \emph{{Large D gravity and charged membrane dynamics
  with nonzero cosmological constant}},
  \href{https://doi.org/10.1007/JHEP12(2018)034}{\emph{JHEP} {\bfseries 12}
  (2018) 034} [\href{https://arxiv.org/abs/1806.08515}{{\ttfamily
  1806.08515}}].

\bibitem{Bhattacharyya:2018iwt}
S.~Bhattacharyya, P.~Biswas and M.~Patra, \emph{{A leading-order comparison
  between fluid-gravity and membrane-gravity dualities}},
  \href{https://doi.org/10.1007/JHEP05(2019)022}{\emph{JHEP} {\bfseries 05}
  (2019) 022} [\href{https://arxiv.org/abs/1807.05058}{{\ttfamily
  1807.05058}}].

\bibitem{Bhattacharyya:2019mbz}
S.~Bhattacharyya, P.~Biswas, A.~Dinda and M.~Patra, \emph{{Fluid-gravity and
  membrane-gravity dualities - Comparison at subleading orders}},
  \href{https://doi.org/10.1007/JHEP05(2019)054}{\emph{JHEP} {\bfseries 05}
  (2019) 054} [\href{https://arxiv.org/abs/1902.00854}{{\ttfamily
  1902.00854}}].

\bibitem{Kar:2019kyz}
A.~Kar, T.~Mandal and A.~Saha, \emph{{The large $D$ membrane paradigm for
  general four-derivative theory of gravity with a cosmological constant}},
  \href{https://doi.org/10.1007/JHEP08(2019)078}{\emph{JHEP} {\bfseries 08}
  (2019) 078} [\href{https://arxiv.org/abs/1904.08273}{{\ttfamily
  1904.08273}}].

\bibitem{Dandekar:2019hyc}
Y.~Dandekar and A.~Saha, \emph{{Large D membrane for Higher Derivative Gravity
  and Black Hole Second Law}},
  \href{https://doi.org/10.1007/JHEP02(2020)083}{\emph{JHEP} {\bfseries 02}
  (2020) 083} [\href{https://arxiv.org/abs/1910.10964}{{\ttfamily
  1910.10964}}].

\bibitem{Biswas:2019xip}
P.~Biswas, \emph{{Stress Tensor for Large-$D$ membrane at subleading orders}},
  \href{https://arxiv.org/abs/1912.00476}{{\ttfamily 1912.00476}}.

\bibitem{Patra:2019hlq}
M.~Patra, \emph{{Comparison between fluid-gravity and membrane-gravity
  dualities for Einstein-Maxwell system}},
  \href{https://arxiv.org/abs/1912.09402}{{\ttfamily 1912.09402}}.

\bibitem{Gregory:1993vy}
R.~Gregory and R.~Laflamme, \emph{{Black strings and p-branes are unstable}},
  \href{https://doi.org/10.1103/PhysRevLett.70.2837}{\emph{Phys. Rev. Lett.}
  {\bfseries 70} (1993) 2837}
  [\href{https://arxiv.org/abs/hep-th/9301052}{{\ttfamily hep-th/9301052}}].

\bibitem{Gregory:1994bj}
R.~Gregory and R.~Laflamme, \emph{{The Instability of charged black strings and
  p-branes}}, \href{https://doi.org/10.1016/0550-3213(94)90206-2}{\emph{Nucl.
  Phys. B} {\bfseries 428} (1994) 399}
  [\href{https://arxiv.org/abs/hep-th/9404071}{{\ttfamily hep-th/9404071}}].

\bibitem{Elvang:2006dd}
H.~Elvang, R.~Emparan and A.~Virmani, \emph{{Dynamics and stability of black
  rings}}, \href{https://doi.org/10.1088/1126-6708/2006/12/074}{\emph{JHEP}
  {\bfseries 12} (2006) 074}
  [\href{https://arxiv.org/abs/hep-th/0608076}{{\ttfamily hep-th/0608076}}].

\bibitem{Horowitz:2001cz}
G.~T. Horowitz and K.~Maeda, \emph{{Fate of the black string instability}},
  \href{https://doi.org/10.1103/PhysRevLett.87.131301}{\emph{Phys. Rev. Lett.}
  {\bfseries 87} (2001) 131301}
  [\href{https://arxiv.org/abs/hep-th/0105111}{{\ttfamily hep-th/0105111}}].

\bibitem{Figueras:2015hkb}
P.~Figueras, M.~Kunesch and S.~Tunyasuvunakool, \emph{{End Point of Black Ring
  Instabilities and the Weak Cosmic Censorship Conjecture}},
  \href{https://doi.org/10.1103/PhysRevLett.116.071102}{\emph{Phys. Rev. Lett.}
  {\bfseries 116} (2016) 071102}
  [\href{https://arxiv.org/abs/1512.04532}{{\ttfamily 1512.04532}}].

\bibitem{Elvang:2007rd}
H.~Elvang and P.~Figueras, \emph{{Black Saturn}},
  \href{https://doi.org/10.1088/1126-6708/2007/05/050}{\emph{JHEP} {\bfseries
  05} (2007) 050} [\href{https://arxiv.org/abs/hep-th/0701035}{{\ttfamily
  hep-th/0701035}}].

\bibitem{Elvang:2007hg}
H.~Elvang, R.~Emparan and P.~Figueras, \emph{{Phases of five-dimensional black
  holes}}, \href{https://doi.org/10.1088/1126-6708/2007/05/056}{\emph{JHEP}
  {\bfseries 05} (2007) 056}
  [\href{https://arxiv.org/abs/hep-th/0702111}{{\ttfamily hep-th/0702111}}].

\bibitem{Elvang:2007hs}
H.~Elvang and M.~J. Rodriguez, \emph{{Bicycling Black Rings}},
  \href{https://doi.org/10.1088/1126-6708/2008/04/045}{\emph{JHEP} {\bfseries
  04} (2008) 045} [\href{https://arxiv.org/abs/0712.2425}{{\ttfamily
  0712.2425}}].

\end{thebibliography}\endgroup

\end{document}